\documentclass{article}
\usepackage[utf8]{inputenc}

\usepackage[margin=2cm]{geometry}
\usepackage[utf8]{inputenc}
\usepackage{url}
\usepackage{array}
\usepackage{amsmath,amssymb,amsfonts}
\usepackage{mathtools}
\usepackage{algorithm}
\usepackage{algpseudocode}
\usepackage{tikz}
\usepackage{graphicx}
\usepackage{subcaption}
\usepackage[export]{adjustbox}
\usepackage{bm}
\usepackage{xcolor}
\usepackage{tabularx}
\usepackage{multirow}
\usepackage[square,sort,comma,numbers]{natbib}
\usepackage{dsfont}
\usepackage{hyperref}

\newcommand{\data}{\pmb{x}}
\newcommand{\threshold}{\pmb{\epsilon}}
\newcommand{\dataObs}{\pmb{x^0}}
\newcommand{\datasim}{\pmb{x^{\mbox{sim}}}}
\newcommand{\distance}{d}
\newcommand{\Model}{\mathcal{M}}

\newcommand{\parameter}{\pmb{\theta}}
\newcommand{\prior}{\pi}

\newcommand{\sac}{\mathcal{S}_{agg-clust}}
\newcommand{\nac}{\mathbb{N}_{agg-clust}}
\newcommand{\np}{\mathbb{N}_{platelet}}
\newcommand{\nacp}{\mathbb{N}_{act-platelet}}
\newcommand{\Pg}{p_{Ag}}
\newcommand{\Pad}{p_{Ad}}
\newcommand{\Pt}{p_{T}}
\newcommand{\Pf}{p_{F}}
\newcommand{\Ra}{a_{T}}

\newcommand{\vap}{v^{AP}_{z}}
\newcommand{\vnap}{v^{NAP}_{z}}

\def\ds{\Delta S}

\newcommand{\statistics}{s}
\newcommand{\R}{\mathbb{R}}
\newcommand{\E}{\mathbb{E}}

\begin{document}

\title{Personalized pathology test for Cardio-vascular disease: Approximate Bayesian computation with discriminative summary statistics learning}

\author{Ritabrata Dutta$^1$\thanks{Corresponding author: Ritabrata.Dutta@warwick.ac.uk}, Karim Zouaoui-Boudjeltia$^2$, 
Christos Kotsalos$^3$,\\
Alexandre Rousseau$^2$, 
Daniel Ribeiro de Sousa$^2$, 
Jean-Marc Desmet$^2$, 
Alain Van Meerhaeghe$^2$,\\
Antonietta Mira$^4$, Bastien Chopard$^3$\\
{\em $^1$University of Warwick, UK}\\
{\em $^2$ Universit{\'e} Libre de Bruxelles, CHU de Charleroi, Charleroi, Belgium}\\
{\em $^3$University of Geneva, Switzerland}\\
{\em $^4$Universit{\'a} della svizzera italiana, Switzerland}\\
}

\date{\today}

\maketitle

\abstract{
Cardio/cerebrovascular diseases (CVD) have become one of the major health issue in our societies. But recent studies show that the present pathology tests to detect CVD are ineffectual as they do not consider different stages of platelet activation or the molecular dynamics involved in platelet interactions and are incapable to consider inter-individual variability. 
Here we propose a stochastic platelet deposition model and an inferential scheme to estimate the biologically meaningful model parameters using approximate Bayesian computation with a summary statistic that maximally discriminates between different types of patients. Inferred parameters from data collected on healthy volunteers and different patient types help us to identify specific biological parameters and hence biological reasoning behind the dysfunction for each type of patients. This work opens up an unprecedented opportunity of personalized pathology test for CVD detection and medical treatment.
}

\section{Introduction}
Cardio/cerebrovascular diseases (CVD) were the first cause of mortality worldwide in 2015, causing 31\% of deaths according to World Health Organization~\cite{WHO}. 
Pathology tests for detection of CVD 
rely on testing functionality of blood platelets, which play a key role in the occurrence of these cardio/cerebrovascular accidents in addition to complex process of blood coagulation, involving adhesion, aggregation on the vascular wall to stop a hemorrhage while avoiding the vessel occlusion. 
In a comprehensive bio-medical evaluation study \cite{breet:2010}, the correlation between the clinical biological measures using platelet function tests and the occurrence of a cardiovascular event was found to be null for half of the techniques and rather modest for others, indicating the evident need for a more efficient tool or method 
to monitor patient platelet functionalities. 
The inadequacy of these tests can be explained by the fact that no current test allows for the analysis of the different stages of platelet activation or the prediction of the in-vivo behavior of those platelets \cite{picker:11,koltai2017platelet}. 
In addition, the current clinical tests do not take into account the dynamic aspect of the process of platelet aggregation and the role that red blood cells can have in this process. 
To address these issues, we extend here the  stochastic model proposed in \cite{chopard2017physical} that simulates numerically the deposition pattern of platelets observed in the Impact-R device~\cite{shenkman:08}, namely the sizes and number of aggregates as a function of time for a layer of whole blood subject to a controlled shear rate.
This model is characterised by bio-physically meaningful parameters the adhesion rate $\Pad$, the aggregation rates $\Pg$ (of depositing on an already deposited platelet) and $\Pt$ (of depositing on an existing cluster of platelets), the deposition rate of albumin $\Pf$, the attenuation factor $\Ra$, and the flux  of activated (AP) and non-activated platelets (NAP) obtained from their characteristics velocities ($\vap$ and $\vnap$). The value of these parameters can be inferred by matching the simulation output with the corresponding in-vitro deposition pattern. 

Our \textbf{\textit{main claim}} here is that the values of some of these model parameters (eg. adhesion and aggregation rates) are precisely the information needed to assess various possible pathological conditions and to quantify their severity with reference to CVD. 

To support
this claim, we develop a methodology to identify medically interpretable parameters differentiating between patients and healthy volunteers. To infer the estimates
of the biologically interpretable parameters of the stochastic platelet deposition model from the deposition patterns observed in the Impact-R device of platelet collected for a patient, we use approximate Bayesian computation (ABC) \cite{lintusaari2017fundamentals, dutta2018parameter} and report the estimated mode of the approximate posterior learned by ABC as the estimates, which can also be interpreted as approximate maximum likelihood estimates (MLE) when non-informative uniform priors are considered on the parameters. We note that ABC inferential algorithms depend on the choice of the summary statistics extracted from the datasets \cite{lintusaari2017fundamentals}, which can be inferred via metric learning \cite{suarez2018tutorial, weinberger2009distance} - a methodology that can provide the summary statistics able to maximally discriminate different patient groups. Leveraging on this crucial link between ABC and metric learning \cite{pacchiardi2020distance}, we are able to identify medically meaningful parameters, which can distinguish between different types of patients and, at the same time, to estimate those parameters for each patient with the aim of developing a test for the pathology. We further notice that the proposed approach can be applied on each patient, in a systematic way. This reduces the bias of a human operator. 

Finally, to verify our proposed methodology, we perform a four stage experiment: \textbf{1)} Collect blood or platelet from 32 patients (16 patients needing dialysis and 16 patients with Chronic Obstructive Pulmonary Disease - COPD) and 16 healthy volunteers; \textbf{2)} Study the deposition patterns observed in the Impact-R of platelet collected for each patient; \textbf{3)} Learn the summary statistics from this dataset which is able to maximally distinguish between the 3 types of patients; \textbf{4)} Estimate the model parameters for each of the patients, using ABC with the use of the learned discriminatory summary statistics from Stage 3. Studying the inferred parameters from each of the patients, we were able to identify medically meaningful parameters which are able to distinguish between patients of different types. 

This study relies only on a small data set and is meant as a proof of concept. A forthcoming clinical study will provide a much larger data set on which we plan to demonstrate the potential of our approach convincingly.

\section{Results and Discussion}
\label{sec:exp_result}
\paragraph{Dataset:} The collected dataset (characterizing the deposition pattern in the Impact-R) can be divided into three groups: healthy volunteers (Group 1), patients needing dialysis (Group 2) and patients affected by Chronic Obstructive Pulmonary Disease (COPD) (Group 3). We collected blood samples from 16 volunteers or patients from each of the three groups. Half of the 16 patients having dialysis were also affected by diabetes and all the volunteers and patients were chosen from a broad age group. 

\begin{figure*}[hbt!]
\caption{{\bf Posterior distribution of the model parameters for a COPD patient.} The green cross indicates the maximum a posteriori (MAP) estimate of the parameters $\Pad \ [s^{-1}]$, $\Pg \ [s^{-1}]$, $\Pt \ [s^{-1}]$, $\Pf \ [s^{-1}]$, $a_T \ [\mu m^2s^{-1}]$, $\vap \ [ms^{-1}]$, $\vnap \ [ms^{-1}]$.}
\centering
\includegraphics[width=.9\textwidth]{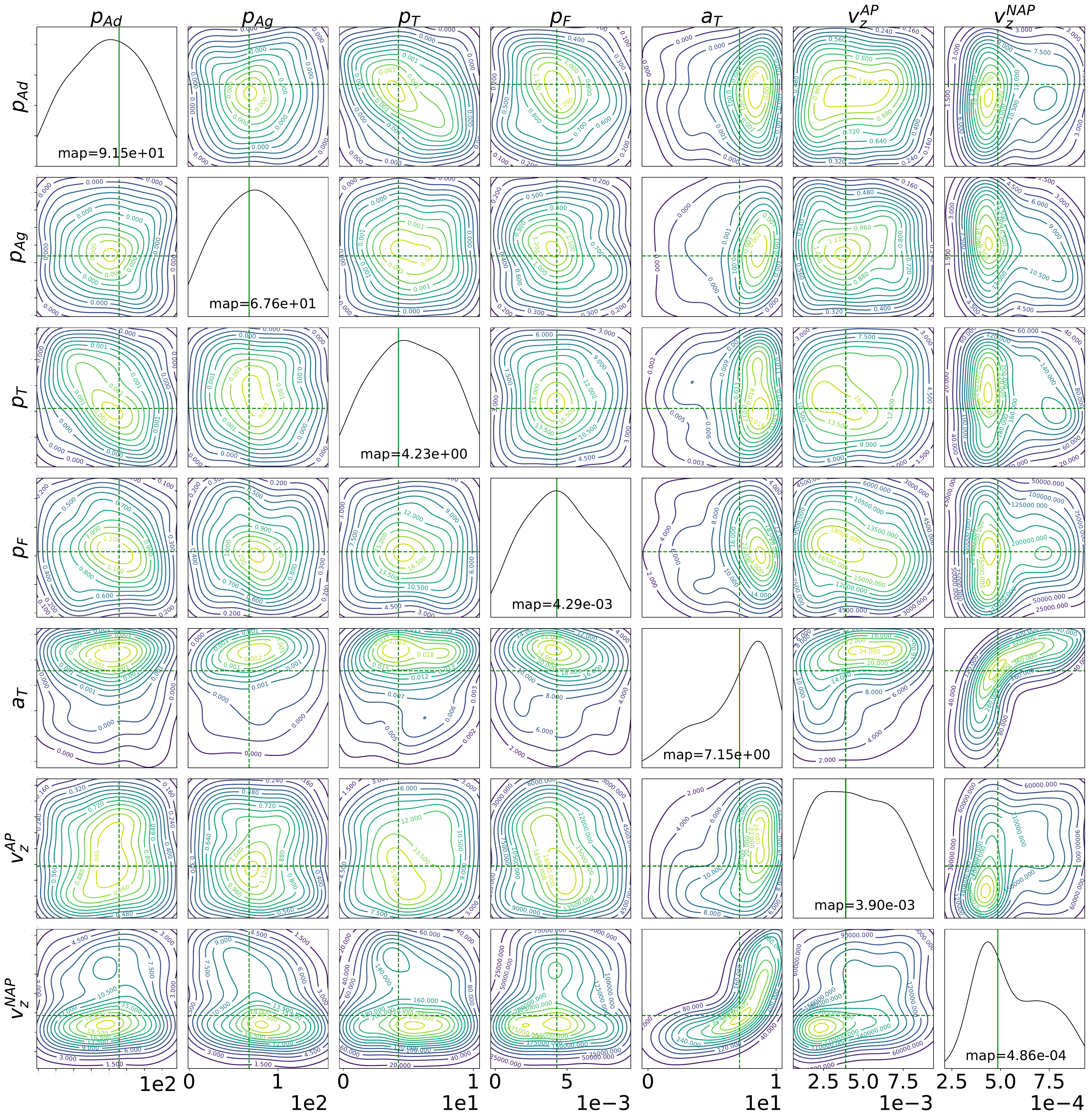}
\label{fig:post_patient34}
\end{figure*}

\paragraph{Summary Statistics learning and Inference of parameters:}
Using Large Margin Nearest Neighbor Metric Learning (LMNN) \cite{weinberger2009distance}, we first learn a 2-dimensional projection of the collected dataset, which maximally discriminates between different types of patients. 
We notice that the two features defining this two-dimensional space do not have a meaningful biological representation as they are just weighted non-linear combinations of the observed time-series of the platelet deposition pattern. We use these two features as summary statistics for ABC and thus the fact that they have no biological interpretation is not relevant since they are only used to facilitate the estimation process of biologically meaningful model parameters that are, in turn, used to define the clinical tests. Next we use the Euclidean distance on this projection space for ABC to infer the parameters of the stochastic platelets depositions model for each of the volunteers (or patients), using the corresponding deposition and aggregation pattern of the platelets in their blood displayed in Impact-R machine. This provides us with a posterior distribution of the parameters given the data from each individual volunteer (patient). After learning the posterior distribution, to provide an estimate of the parameters for each of the patients, we calculate the maximum a posteriori (MAP) estimate of the parameters. Here we note that the Bayesian point estimates are minimizers of posterior loss (eg. posterior mean minimizing squared error loss or MAP minimizing 0-1 loss) \cite{berger2013statistical}. These ABC inferred parameters provide valuable biological interpretation. In Figure~\ref{fig:post_patient34}, we illustrate the inferred posterior distribution for a patient with COPD and the corresponding maximum a posteriori (MAP) estimate of the parameters. 

\begin{figure*}[hbt!]
\caption{{\bf Boxplot of the MAP estimates of the biologically meaningful discriminating parameters grouped according to patient types.} Healthy: 
Healthy volunteers (16 patients); Dialysis: All Patients undergoing dialysis (16 patients);  COPD: Patients with COPD (16 patients). The box in the plot extends from the first quartile (Q1) to third quartile (Q3) of the estimated parameters for each type of patients, with an orange horizontal line at the median. The upper whisker extends to the largest value less than Q3 + 1.5 * IQR and the lower whisker extends to the lowest value greater than Q1 - 1.5 * IQR, where IQR is the interquartile range (Q3-Q1). 
Beyond the whiskers, any values are considered outliers and are plotted as individual points. }
\centering
		\begin{subfigure}{.32\linewidth}
			\centering
			\includegraphics[width=\linewidth]{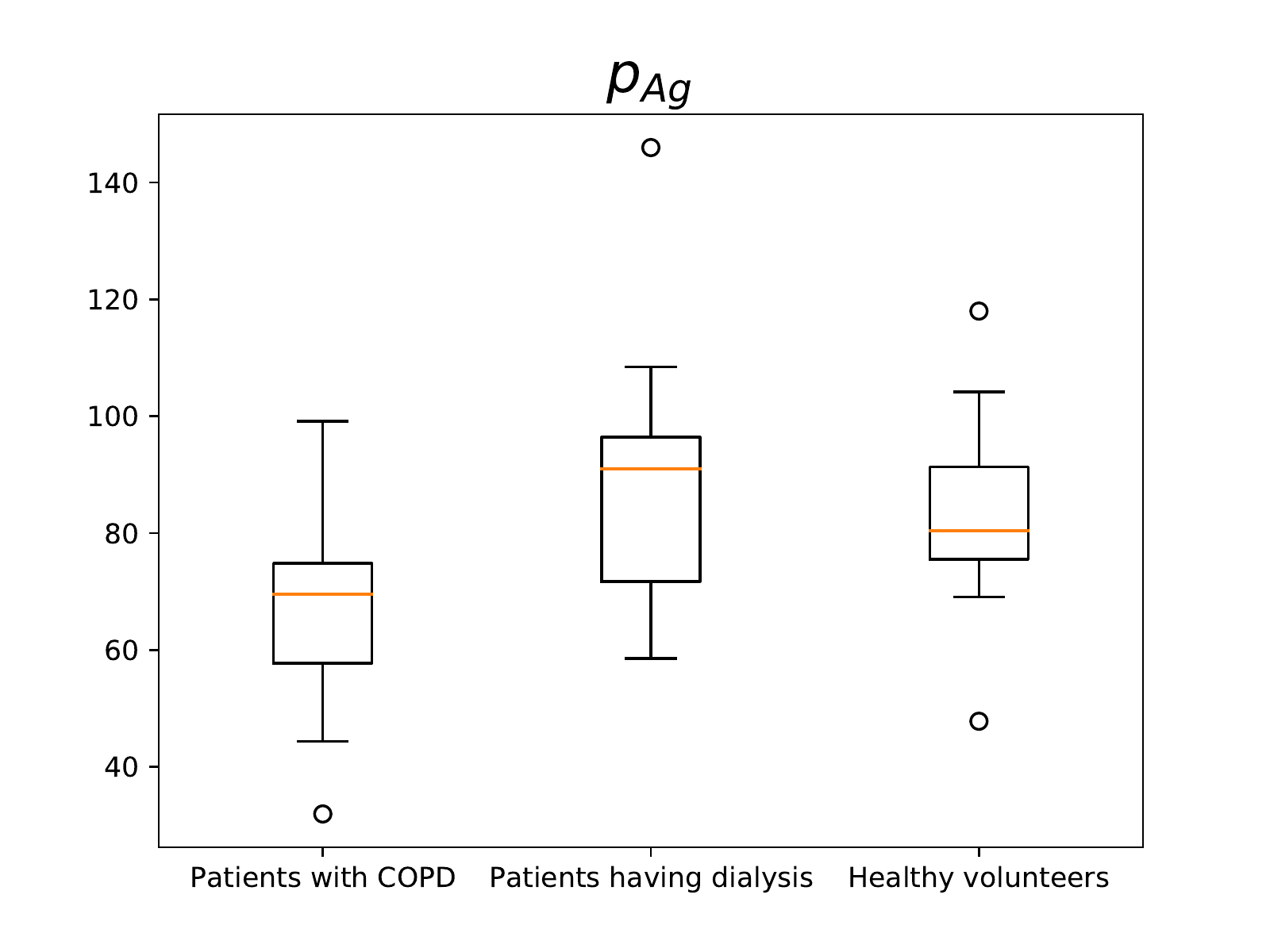}
			\caption{$p_{Ag}$}
		\end{subfigure}
		\begin{subfigure}{.32\linewidth}
			\centering
			\includegraphics[width=\linewidth]{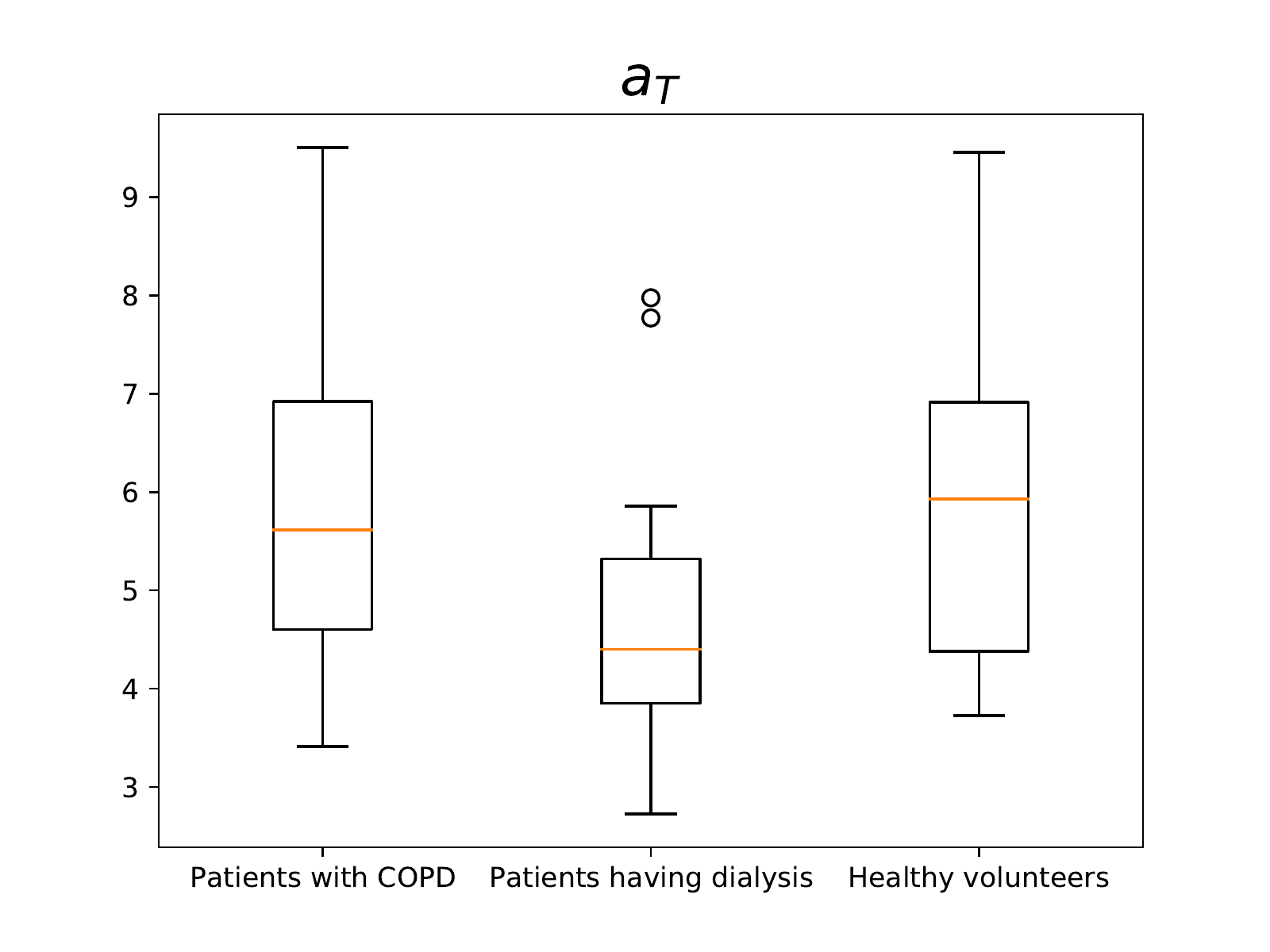}
		\caption{$a_{T}$}
		\end{subfigure}
		\begin{subfigure}{.32\linewidth}
			\centering
			\includegraphics[width=\linewidth]{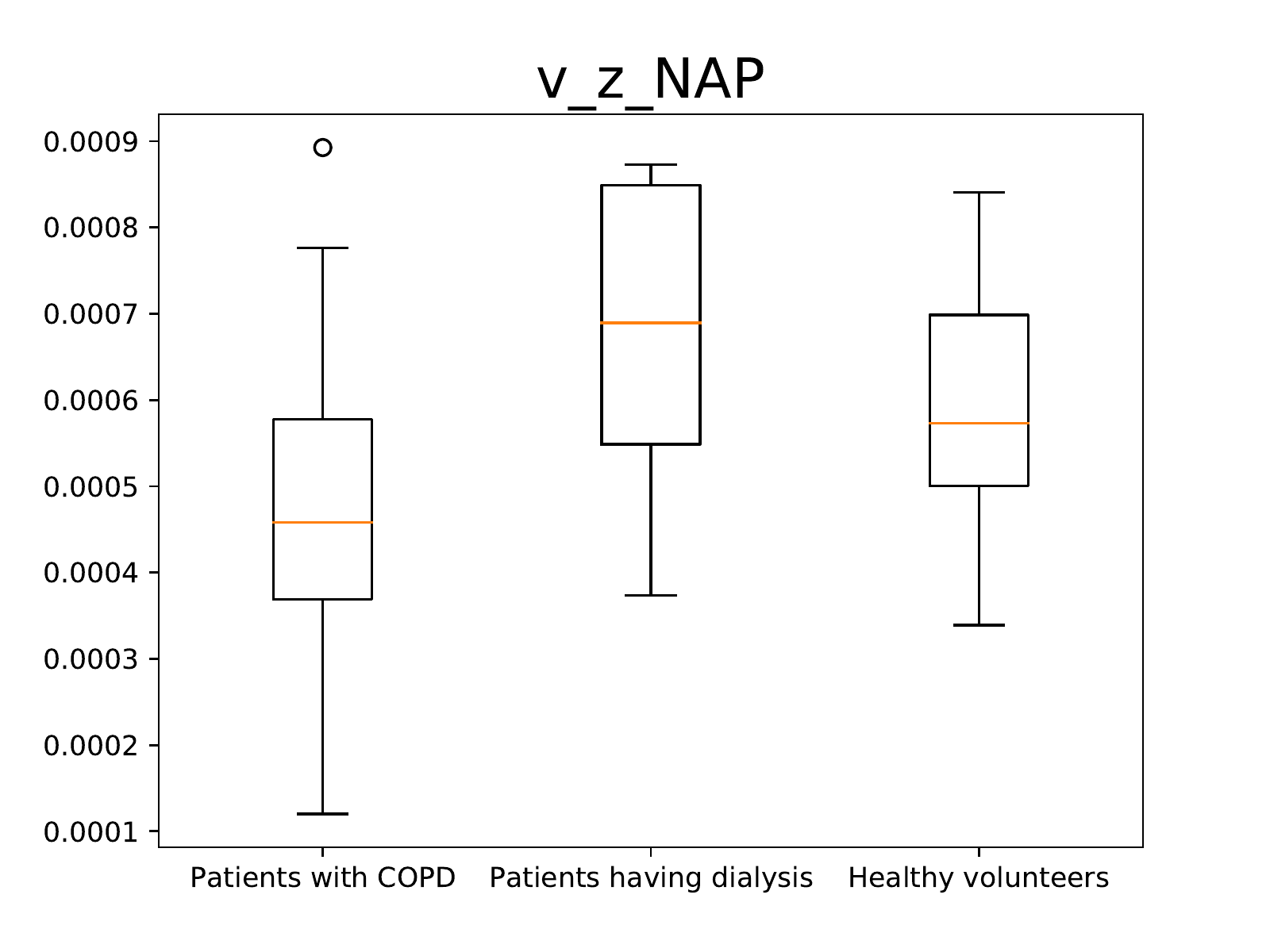}
		\caption{$v_z^{NAP}$}
		\end{subfigure}
\label{fig:boxplot_estimate}
\end{figure*}   

\paragraph{Uncertainty in MAP estimation:} We illustrate the uncertainty of the MAP estimates in each patient group through the boxplots of MAP estimates for each patient in each of the three groups in Figure~\ref{fig:boxplot_estimate} for the three most discriminative parameters. The average observed standard deviation across the three groups are $\sigma^*_{\Pg}:17.79, \sigma^*_{a_T}:1.64, \sigma^*_{v_z^{NAP}}:1.6e-04$. 

One of our main assumption to build the test for pathology is that 
the parameter values for each of the patients in a specific group are centered around a true value of the parameters with a small variance. 
Under this assumption, using consistency theorem for MAP estimates \cite{van2000asymptotic} we can argue that the MAP estimates for each patient in that group would converge to the true value if we have increasingly many samples (eg. repeated measurements using Impact-R machine) for each of the patients. Given we only have one sample from each of the patient, we can argue that the distribution of the MAP estimates of the patients in a group would be centered around the true value. To justify the standard deviation observed in the MAP estimates of the patients in a group, we simulated 10 dataset using a true parameter value and computed the standard deviation of the MAP estimates for each of the simulated dataset ($\hat{\sigma}_{\Pg}:23.4, \hat{\sigma}_{a_T}:0.95, \hat{\sigma}_{v_z^{NAP}}:2.71e-05$). We notice that the variance in the MAP estimates of the real data in a group [Figure~\ref{fig:boxplot_estimate}] is similar, in its order of magnitude, to the one observed in our simulation study .

Finally, we notice that the mean/median of the MAP estimates in each group of patients are significantly different among the three groups for the parameters $\Pg, a_T, v_z^{NAP}$, possibly indicating a discriminative behaviour among the three groups, which helped us to identify the most discriminative parameters and devise a test for pathology. 

\begin{table*}[hbt!]
\centering
    \caption{{\bf Statistics (P-values, P-values corrected for multiple testing) of Kruskal-Wallis test using maximum a posteriori (MAP) estimates of the parameters.} A higher value of the statistics and P-values smaller than 0.05 (bold) indicates that the medians of the estimated parameters of the corresponding groups are statistically different with a significance cutoff of 0.05. Healthy indicates healthy volunteers, Dialysis indicates patients under dialysis, and COPD stands for  patients with COPD. The correction of P-values for multiple testing has been done using the Benjamini-Hochberg procedure \cite{benjamini1995controlling}.}
    \label{tab:MAP_Post_mean_KW}
    \begin{tabular}{|c|c|c|c|c|}
    \hline
    Parameters & All three classes & Healthy vs Dialysis & Healthy vs COPD & Dialysis vs COPD \\\hline
$\Pad$ & 2.7 (2.5e-1, 3.5e-1) & 2.2 (1.3e-1, 2.3e-1) & 3.5e-2 (8.5e-1, 9.1e-1) & 1.7 (1.8e-1, 2.4e-1)\\
$\Pg$ & \textbf{9.3 (9.5e-3, 4.9e-2)} & 2.4e-1 (6.2e-1, 7.9e-1) & \textbf{7.3 (6.6e-3, 4.6e-2)} & \textbf{6.3 (1.1e-2, 4e-2)}\\
$\Pt$ & 2.3e-2 (9.8e-1, 9.8e-1) & 1.4e-3 (9.7e-1, 9.7e-1) & 2.2e-2 (8.8e-1, 9.1e-1) & 1.3e-2 (9.1e-1, 9.1e-1)\\
$\Pf$ & 1.4 (4.9e-1, 5.7e-1) & 1.7e-1 (6.7e-1, 7.9e-1) & 4.1e-1 (5.2e-1, 9.1e-1) & 1.5 (2.1e-1, 2.4e-1)\\
$\Ra$ & 4.9 (8.2e-2, 1.9e-1) & \textbf{3.8 (5.0e-2, 2.2e-1)} & 1.2e-2 (9.1e-1, 9.1e-1) & 3.5 (5.9e-2, 1.3e-1)\\
$\vap$ & 3.7 (1.5e-1, 2.7e-1) & 2.7 (9.7e-2, 2.2e-1) & 3.5e-2 (8.5e-1, 9.1e-1) & 2.7 (9.7e-2, 1.7e-1)\\
$\vnap$ & \textbf{8.5 (1.4e-2, 4.9e-2)} & 2.8 (8.9e-2, 2.2e-1) & 2.5 (1.1e-1, 3.9e-1) & \textbf{7.3 (6.6e-3,4e-2)}\\    \hline
    \end{tabular}
\end{table*}

\paragraph{Kruskal-Wallis H-test:}Our main claim is based on the assumption that the median of the MAP estimates of the three patient groups are different among the groups for some of the parameters. We notice that the distribution of the MAP estimates of patients from different groups can still overlap due to the uncertainty in MAP estimation. To test whether the patient specific estimated parameters of the model can distinguish patients from healthy volunteers and furthermore discriminate between different types of the patients, we use the Kruskal-Wallis H-test \cite{kruskal1952use} which tests \textit{the null hypothesis that the median of the MAP estimates of the different types of patients are equal versus a bilateral alternative.}
The computed test statistics and P-values between all the three groups (healthy volunteers, patients needing dialysis and patients with COPD) are reported in the first column of Table~\ref{tab:MAP_Post_mean_KW}. Considering a cutoff for significance of $0.05$, the null hypothesis gets rejected for the parameters $\Pg$ and $\vnap$
indicating that the values of these parameters significantly differ between the groups. 

We notice that the rejection of the null hypothesis does not indicate which of these groups differ. Hence, we perform post-hoc Kruskal-Wallis H-test between healthy volunteers and patients needing dialysis, healthy volunteers and patients with COPD and between patients needing dialysis and patients with COPD  [columns 2, 3 and 4 of Table~\ref{tab:MAP_Post_mean_KW} respectively]. This indicates that $\Pg$ can clearly differentiate patients with COPD from both healthy volunteers and patients having dialysis. Further $\Ra$ and $\vnap$ being able to discriminate patients needing dialysis correspondingly from healthy volunteers and patients with COPD. We list these parameters which are capable to distinguish between the corresponding groups in Table~\ref{tab:disc_param}.
\begin{table}[hbt!]
\centering

\caption{{\bf Biologically meaningful discriminating parameters.} The parameters which are significantly different between the corresponding two groups of patients (volunteers).}
    \label{tab:disc_param}
    \begin{tabular}{|c|c|}
    \hline
     & Discriminating parameters \\\hline
    Healthy vs Dialysis & $\Ra$  \\\hline
    Healthy vs COPD & $\Pg$ \\\hline
    Dialysis vs COPD & $\Pg$, $\vnap$  \\\hline
    \end{tabular}
\end{table}

\paragraph{Discriminating parameters and pathology test:} According to our analysis, the biologically meaningful parameters which are able to discriminate between different patient types up to some accuracy are $\Pg$, $\Ra$ and $\vnap$, as shown in [Table~\ref{tab:disc_param} and Figure~\ref{fig:boxplot_estimate}]. Further, these parameters can be divided into two distinct types of pathologies related to biochemical and conformational changes  correspondingly in platelets and red blood cells (RBCs). 
 
\textit{Pathological changes in platelets:} The first group of parameters ($\Pg$, $\Ra$) represent explicit intrinsic changes in platelets associated with the presence of pathology, which causes a change in their patterns of adhesion and aggregation. The common thread between dialysis patients and COPD patients is the existence of chronic systemic inflammation implicated in the development of cardiovascular disease. In response to inflammation, it is well known that platelets in COPD and dialysis patients are activated in the bloodstream, altering their hemostatic properties \cite{mourikis2020platelet, thijs2008mild, mallah2020platelets, malerba2016platelet} and therefore the process of adhesion and aggregation. 

\textit{Biochemical and conformational changes in RBCs:} $\vnap$ reflects another aspect of the presence of pathology, where changes in the velocity of platelets are caused via their interaction with red blood cells (RBCs), which may have undergone biochemical and conformational changes altering blood rheology \cite{bujak2015prognostic} under pathological situations. These changes in RBCs are known as spherization and have been observed in sepsis, dialysis patients, COPD patients and other pathologies causing chronic or acute systemic inflammation \cite{piagnerelli2007assessment}. Recently, \cite{zouaoui2020spherization} reported that the RBC spherization induces an increase in platelet adhesion and aggregation processes and an increase in platelet transport to the wall. Hence, the changes observed in platelets velocities under pathologies may indicate medical conditions which causes biochemical and conformational changes to RBCs.

\begin{table}[hbt!]
\centering
\caption{{\bf Sensitivity and Specificity of the proposed test} using the most discriminative parameters to identify diseased patients compared to healthy volunteers.}
    \label{tab:sens_spec_patho_test}
    \begin{tabular}{|c|c|c|}
    \hline
    & Healthy vs COPD & Healthy vs Dialysis \\\hline
    \textit{Discriminating parameter} &$\Pg$& $\Ra$ \\\hline
    Sensitivity & 0.75 & 0.56\\\hline
    Specificity & 0.75 & 0.62\\\hline
    \end{tabular}
\end{table}

\textit{Pathology test:} Based on the most discriminative parameters ($\Ra$ and $\Pg$) identified by our analysis, we devise a test for pathology to identify diseased patients (correspondingly for patients with COPD and patients having dialysis). An individual is identified as healthy or belonging to COPD group (similarly to dialysis patients) if the MAP estimate of the parameter $\Ra$ (correspondingly $\Pg$) is closer to the median of the MAP estimates of the healthy volunteers 
or closer to the corresponding value estimated on patients in the COPD group (respectively dialysis patients). The sensitivity and specificity \cite{shreffler2020diagnostic} of our two tests are reported in the Table~\ref{tab:sens_spec_patho_test}. This shows that we can identify patients having COPD with higher degree of accuracy given that our analysis only depends on a relatively small number of patients/volunteers.

\section{Materials and methods}
\label{sec:matmethod}Based on detailed \emph{in vitro} experiments using the Impact-R device mimicking platelet adhesion-aggregation in blood vessels, first we provide a model which is an \emph{in silico} counterpart for an in-depth description and understanding of the phenomenon and the underlying mechanisms. 

\subsection{Impact-R experiment}
\label{sec:impactR}
\begin{figure}[!ht]
\caption{\textbf{Data from Impact-R device:} The data collected from Impact-R for a healthy volunteer (a, b, c) containing three observed quantities $\nac(t)$, $\sac(t)$ and $\np(t)$ are correspondingly average size of the aggregation clusters, their number per $mm^2$ and the number of non-activated platelets per $\mu\ell$ still in suspension at time $t$.}
	\centering
		\begin{subfigure}{.32\linewidth}
			\centering
			\includegraphics[width=\linewidth]{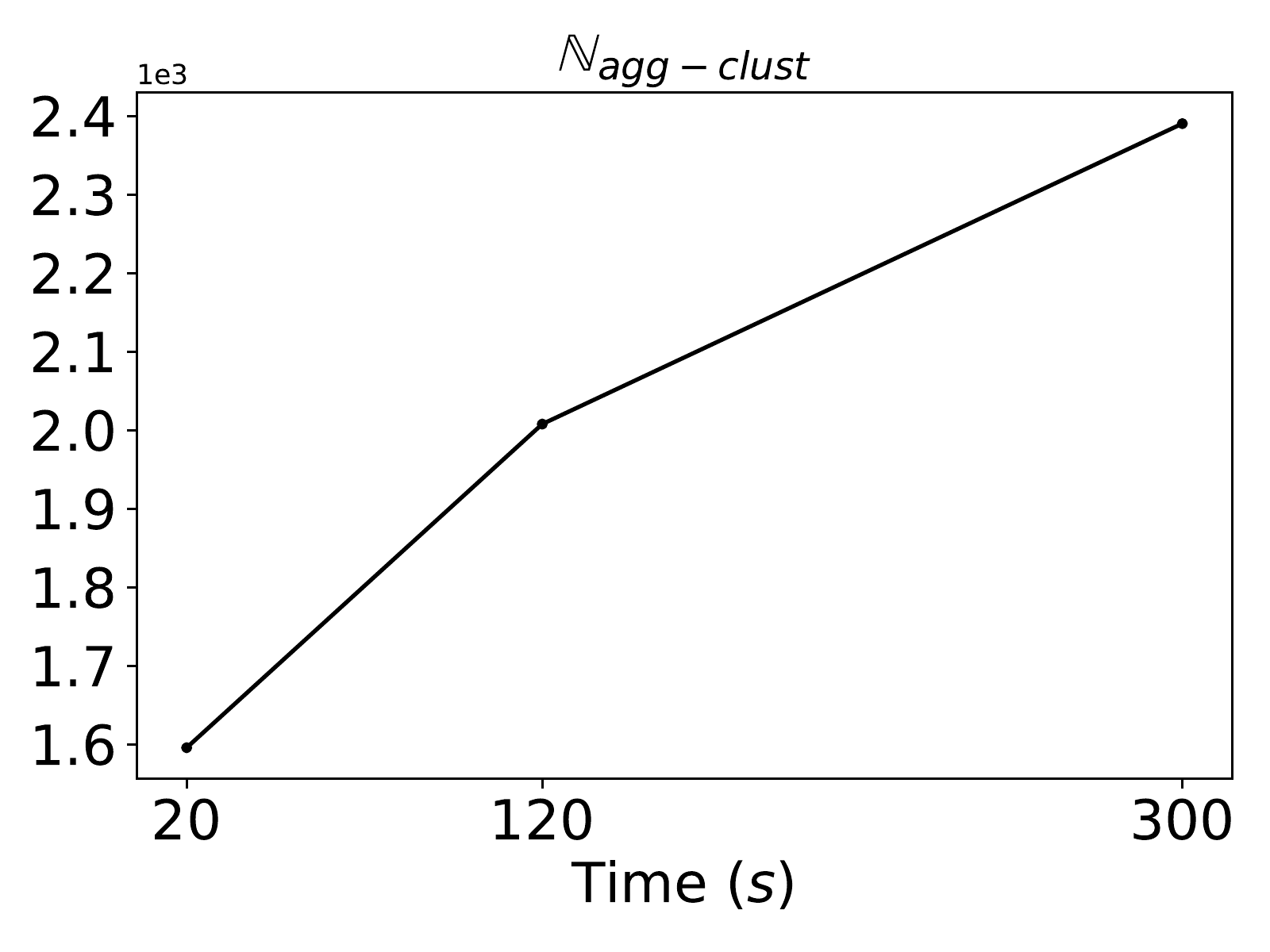}
			\caption{$\nac$}
		\end{subfigure}
		\begin{subfigure}{.32\linewidth}
			\centering
			\includegraphics[width=\linewidth]{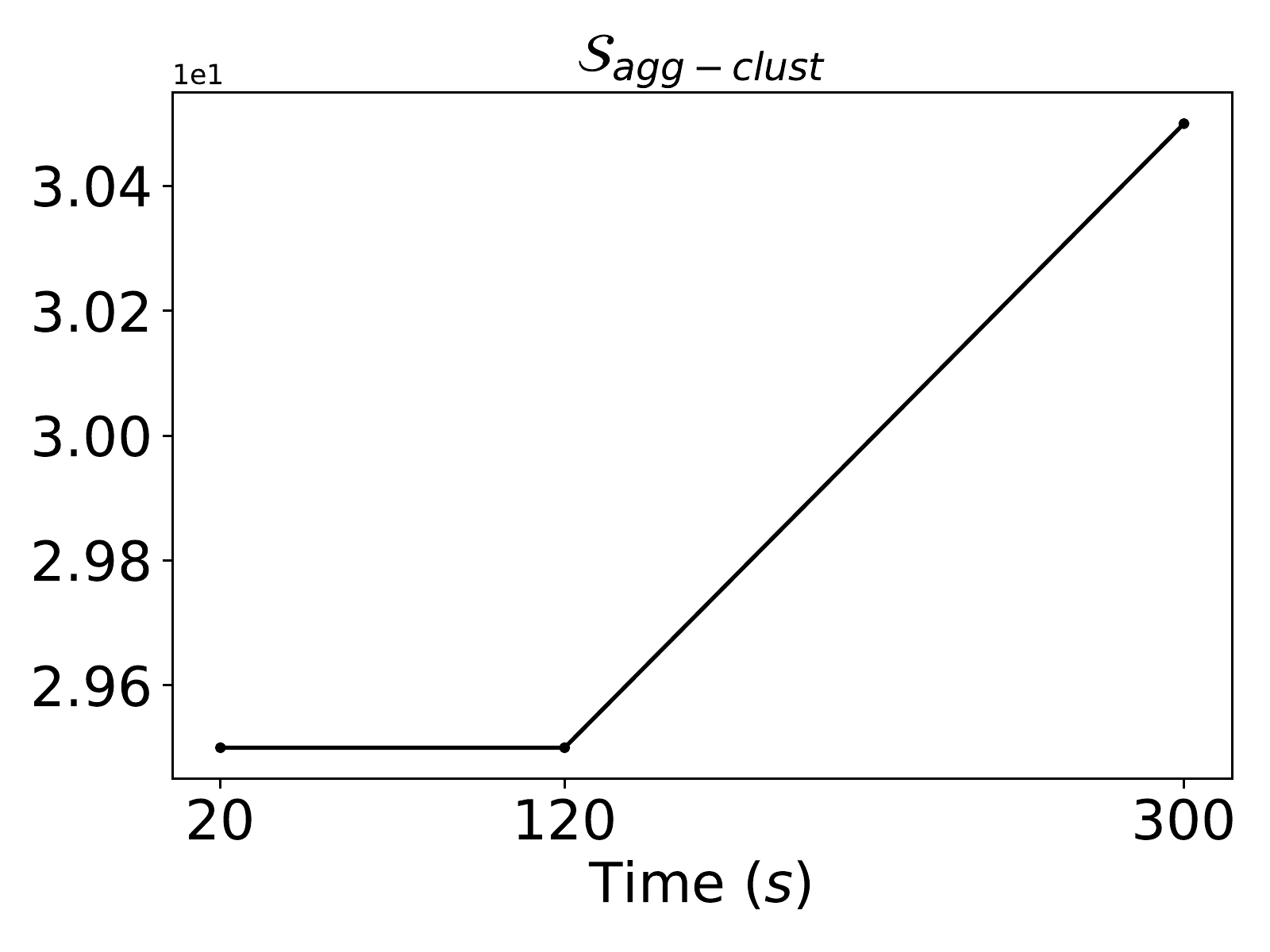}
			\caption{$\sac$}
		\end{subfigure}
		\begin{subfigure}{.32\linewidth}
			\centering
			\includegraphics[width=\linewidth]{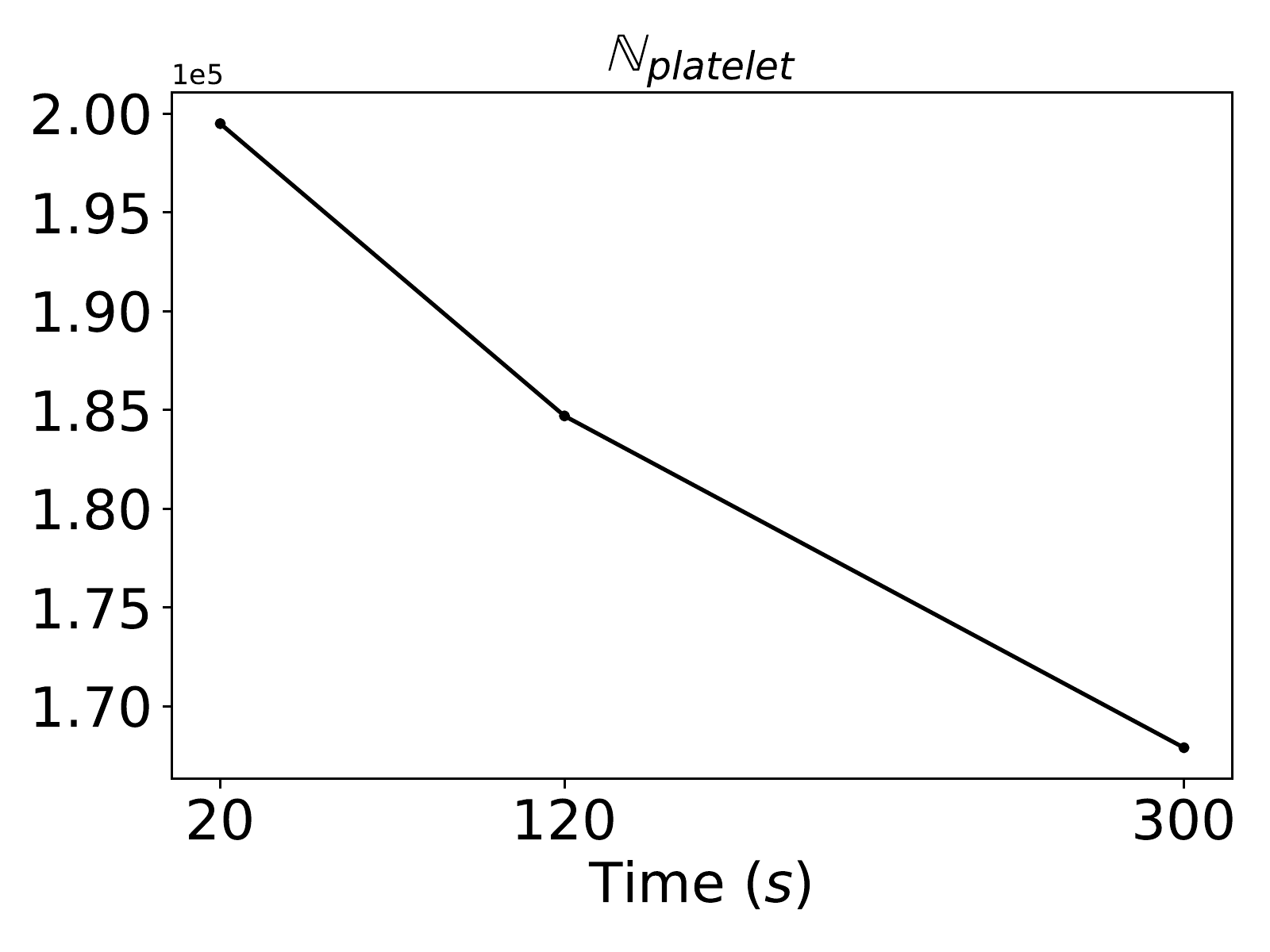}
		\caption{$\np$}
		\end{subfigure}
\label{fig:ImpactR-data-healthyvolunteer}
\end{figure}

Impact-R~\cite{shenkman:08}, a well-known platelet function
analyzer, is a cylindrical device whose bottom wall is a fixed disc (deposition substrate), while the upper wall is a rotating disc (shaped as a cone with a small angle). The height of the device is $0.82~mm$ and due to the motion of the upper wall a pure shear flow is created. A controlled shear rate $\dot{\gamma}$ is produced in a given observation window of $1 \times 1~mm^2$, where we track the formation of clusters resulting from the deposition and aggregation of platelets. Blood was drawn from both healthy and diseased donors with different hematocrit (volume fraction of red blood cells (RBC)). Before starting the tests, a sample is recovered and analysed, to determine the concentration of activated (AP) and non-activated (NAP) platelets. Serum albumin, the most abundant protein in human blood plasma, antagonises with the platelets, preventing them from adhering to the substrate.
The quantities of interest are the number of clusters and their size formed in the substrate, and the number of AP still in suspension. Our goal is to explain the observed (\emph{in vitro} experiments) time evolution of these three quantities, which is illustrated for a healthy volunteer in Figure~\ref{fig:ImpactR-data-healthyvolunteer}.

\subsection{Stochastic model of platelet deposition} 
\label{sec:stochmodel}

\begin{figure}[!ht]
\caption{\textbf{Window of Impact-R device:} The bottom wall is a fixed boundary of dimensions $1 \times 1~mm^2$, the wall-bounded direction is $0.82~mm$. The bulk contains whole blood at different hematocrit. The discretization of the substrate is such that in every cell can fit just one platelet. The initial densities of the blood particles are determined by the \emph{in vitro} experiment and usually are about: $172~200~(\mu l)^{-1}$ for NAP, $4808~(\mu l)^{-1}$ for AP, $2.69 \times 10^{13}~(\mu l)^{-1}$ for Al. The image next to the discretised substrate corresponds to the \emph{in vitro} experiment.}
	\centering
	\includegraphics[width=\linewidth]{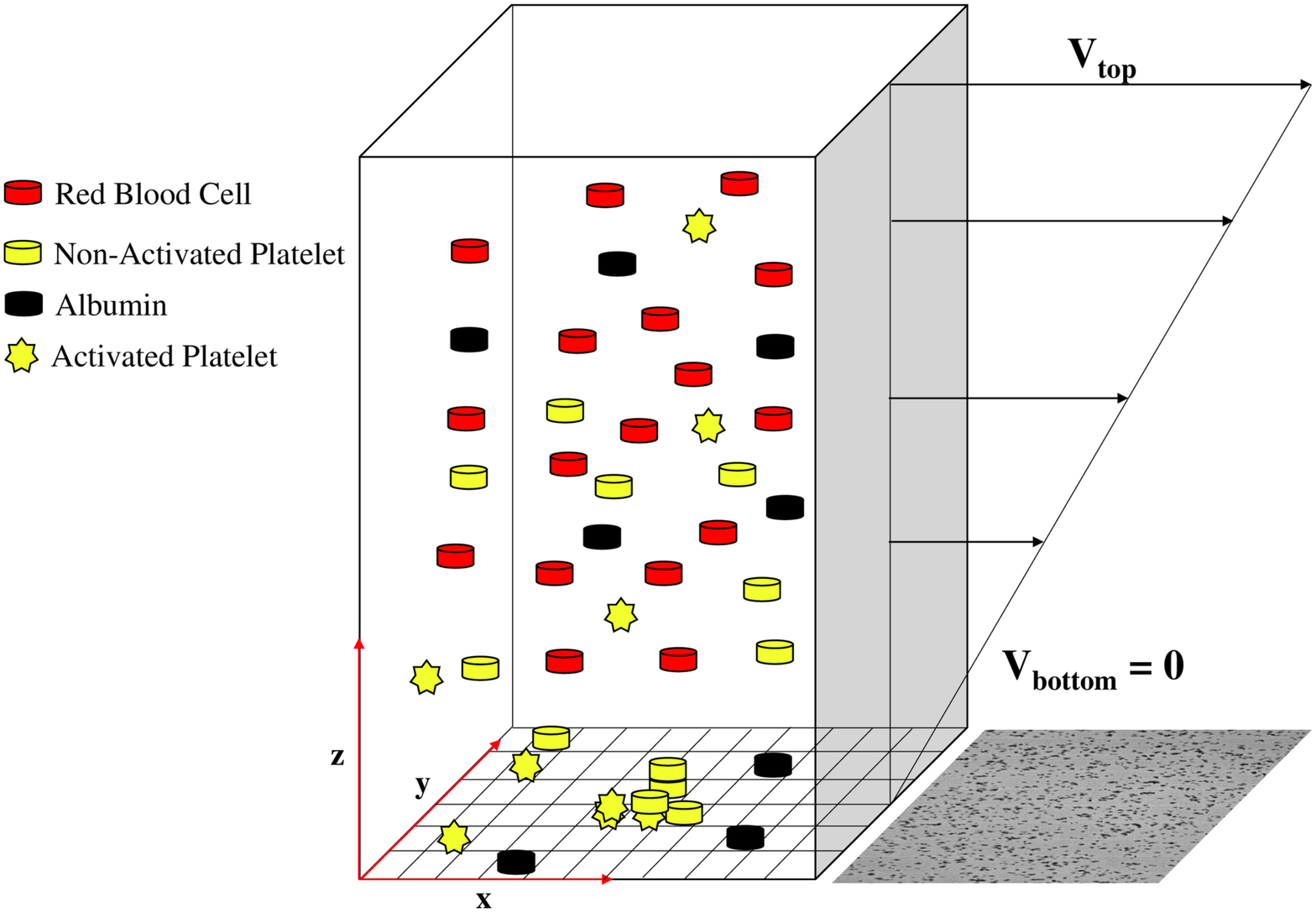}
\label{fig:ImpactR-window}
\end{figure}

The deposition process in Impact-R was successfully described with a
mathematical model for the first time in~\cite{Chopard_2015,Chopard_2017} accounting for the following observations: (i) AP adhere to the deposition surface, forming a seed for a new cluster, (ii) NAP and AP can deposit at the periphery or on top of an existing cluster and (iii) Albumin (Al) deposits on the surface, thus reducing locally the adhesion and aggregation rates of platelets. A sketch of the situation shown in Figure~\ref{fig:ImpactR-window}. In \cite{Chopard_2017}, the platelets reach the bottom layer due to a RBC-enhanced shear-induced 1D diffusion. In the present work, we propose a fully stochastic model of platelet deposition, by substituting the 1D diffusion systems with a 3D random walk, while keeping the deposition dynamics the same. The reason of this new approach is to avoid the introduction of a boundary layer (denoted by $\Delta z$ in \cite{Chopard_2017}) when  coupling  platelet transport with platelet deposition. With this particle based approach, a platelet becomes a candidate for deposition whenever it hits the deposition surface. In \cite{Chopard_2017} the diffusion coefficient $D$ and the thickness of the boundary layer $\Delta z$ where determined independently of the other model parameters, precisely by assuming a random motion of the platelets. Now, we keep the same level of description everywhere and, instead of $D$ and $\Delta z$, two new parameters are considered, namely the characteristic velocities of activated and non-activated platelets. They will be inferred from the data, together  with the adhesion and aggregation rates, and other quantities defined below.

The random walk of the platelets can be described by their jump $(\Delta z (t), \Delta x (t), \Delta y (t))$ at each iteration $t$:
\begin{align}
    \Delta z (t) &= \lambda ~ v_z ~ |s_z| ~ dt, \\
    \Delta x (t) &= v_{xy} ~ |s_{xy}| ~ \cos(2 \pi r) ~ dt, \\
    \Delta y (t) &= v_{xy} ~ |s_{xy}| ~ \sin(2 \pi r) ~ dt,
\end{align}
where $v_{z,xy}$ is a speed unit, $r$ is a random variable uniformly distributed in $[0,1]$, $s_{z,xy} \in ( -\infty, \infty )$ is a random variable distributed as a standard normal distribution, $\lambda \in \{-1,1\}$ with probability $1/2$ for each outcome, and $dt$ is the time step of the simulation. Superimposing the stochastic motion with the velocity field of the pure shear flow, the positions of the platelets are updated as
\begin{align}
    z_i (t+dt) &= z_i (t) + \Delta z_i (t), \\
    x_i (t+dt) &= x_i (t) + \Delta x_i (t) + \dot{\gamma}~z_i~dt, \\
    y_i (t+dt) &= y_i (t) + \Delta y_i (t).
\end{align}

It should be noted that this stochastic model is physically equivalent to the shear-induced diffusion model used in \cite{Chopard_2017}. From the random motion of particles, diffusion constants emerge both along the flow direction and perpendicular to it. The equivalence of random walk and diffusion is a well known result, see for instance \cite{Xin2016}. It has been checked that the platelet deposition pattern obtained with this new model is the same, up to statistical fluctuations, as with the model used in \cite{Chopard_2017}, when the physical properties are the same.
The present model is computationally more costly than that of \cite{Chopard_2017}, but provides more flexibility for parameter inference, in particular to infer, from in vitro data, the transport properties of platelets towards the deposition surface. Additionally the present transport model no longer assumes the same transport properties for activated and non-activated platelets. Actually, this feature allows us to better explain the Impact-R data as activated platelets turn out to move faster, probably due to their increased effective hydrodynamic size. As a consequence, the final results of both models cannot be compared. 

Our interest here is to model the transport of platelets in the direction perpendicular to the blood flow so as to obtain the flux of platelets that will reach the so-called Cell Free Layer (the layer without red blood cells near a wall), at the bottom of the Impact-R device and be candidate for deposition. The stochastic transport along the blood flow direction is not expected to have an impact on the deposition, due to the fast mixing of platelet in the horizontal plan. This assumption made in  \cite{Chopard_2017} is confirmed by the present model which includes explicitly this horizontal transport. Note finally that here we can exclude any drift in the vertical direction due to the up-down symmetry in the Impact-R setting. This absence of a drift, sometimes proposed to describe platelet transport in a tube, has been confirmed by full resolved blood flow simulations, in which deformable red blood cells and platelets are in suspension in a plasma subject to a shear flow \cite{Kotsalos2019DigitalBlood}.

Owing to the different dynamics and physics governing the activated and non-activated platelets, they have in principle different speed units ($v^{AP}_{z,xy}, v^{NAP}_{z,xy}$). Regarding the motion of albumin, its abundance allows us to neglect the small density gradients due to its deposition, and thus albumin can deposit at any time at a maximum value of deposition rate.

The AP and NAP that cross the lower boundary of the computational domain are removed from the bulk if the deposition is successful, or get trapped at the Cell Free Layer (CFL) for a future deposition attempt (they never get re-injected into the bulk). Further, periodic conditions are applied at the $x,y$ directions, and bounce back boundary condition for the platelets that cross the upper boundary. Assuming a good horizontal mixing in the $xy$-plane due to the rotating flow, and given its low impact on the deposition process, the stochastic part of the motion in the $x,y$ directions can be fixed to the same order of magnitude as the $z$ direction velocity. Therefore, we consider $v^{AP}_{xy} = v^{AP}_{z}$, $v^{NAP}_{xy} = v^{NAP}_{z}$ and $s_{xy} = s_{z}$.

Next we describe the deposition rates. 
Let us denote by $N_{i,j}(t)$ the number of candidate particles for deposition above the cell at position $i,j$ of the discretised substrate. 
The deposition of the platelets and albumin on the substrate follow the stochastic rules described in \cite{Chopard_2017}, i.e., based on the $N_{i,j}(t)$ and on the occupancy of the $i,j$-th cell at time $t$. Albumin that reaches
the substrate at time $t$ deposits with a probability $P(t)$ which
depends on the local density $\rho_{al}(t)$ of already deposited
albumin. We assume that $P$ is proportional to the remaining free space
in the cell,
\begin{equation}
  P(t)=\Pf(\rho_{max}-\rho_{al}(t)) dt,
\end{equation}
where $\Pf$ is a parameter to be determined and $\rho_{max}$ is given by the
constraint that at most 100,000 albumin particles can fit in a
deposition cell of area $\ds=5~(\mu m)^2$, corresponding to the size of
a deposited platelet (obtained as the smallest variation of cluster
area observed with the microscope). An activated platelet that hits a platelet-free cell deposits with a probability $Q$, where $Q$ decreases as the local concentration $\rho_{al}$ of albumin increases. We assumed that
\begin{equation}
  Q=\Pad\exp(-\Ra\rho_{al}) dt,
\end{equation}
where $\Pad$ and $\Ra$ are parameters to be determined. This expression can be
justified by the fact that a platelet needs more free space than an
albumin to attach to the substrate, due to their size difference.  In other words, the probability of having enough space for a platelet, decreases roughly exponentially with the density of albumin in the substrate (more details in \cite{chopard2017physical}).  
In our model, AP and NAP can deposit next to already
deposited platelets.  From the above discussion, the aggregation
probability $R$ is assumed to be
\begin{equation}
  R=\Pg\exp(-\Ra\rho_{al}) dt,
\end{equation}
with $\Pg$ another unknown parameter. We also introduce $\Pt$ the rate
at which platelets deposit on top of an existing cluster. 
Figure \ref{fig:ImpactR-window} presents coarsely the competing adhesion-aggregation process between albumin and platelets. More details on the stochastic deposition rules can be found in \cite{Chopard_2017}.

For the purpose of the present study, the platelet deposition model $\Model$ is parametrized in terms of the seven quantities introduced above, namely
the adhesion rate $\Pad$, the aggregation rates $\Pg$ and $\Pt$, the
deposition rate of albumin $\Pf$, the attenuation factor $\Ra$, and the velocities of AP and NAP $\vap$ and $\vnap$. 
Collectively, we define
\[ \parameter= (\Pg, \Pad, \Pt, \Pf, \Ra, \vap, \vnap). \]
If the initial number of AP and NAP at time $t=0$ ($\np(0)$ and $\nacp(0)$), as well as the concentration of albumin are known from the experiment, we can forward simulate the deposition of platelets over time using model $\Model$ for the given values of these parameters $\parameter = \parameter^*$:
\begin{eqnarray*}
\label{eq:simulator_depo}
\Model [\parameter = \parameter^*] \rightarrow 
\left\lbrace 
\left(\nac(t),\sac(t),\np(t)\right), \ t=0, \ldots, T \right\rbrace.
\end{eqnarray*}
where $\nac(t)$,$\sac(t)$ and $\np(t)$ are
correspondingly average size of the aggregation clusters, their number
per $mm^2$, the number of non-activated and pre-activated platelets
per $\mu\ell$ still in suspension at time $t$.

The Impact-R experiments have been repeated with the whole blood obtained from each of the volunteers and patients and the observations were made at time, 20 sec., 120 sec. and 300 sec. At these three time points, $\left(\nac(t), \sac(t),
\np(t) \right)$ are measured [Figure~\ref{fig:ImpactR-data-healthyvolunteer}]. Let us call the observed dataset 
collected through experiment as,
\begin{eqnarray*}
\dataObs \equiv 
\left\lbrace \left(\nac^0(t),\sac^0(t),\np^0(t)\right): t = 0 \mbox{ sec.}, \ldots, 300 \mbox{ sec.}\right \rbrace.
\end{eqnarray*}
\subsection{Estimation of model parameters}
\label{sec:estimation}
As the likelihood function induced by the platelets deposition model is analytically intractable due to the need of computing a very high-dimensional integral, we can not compute the maximum likelihood estimate of the model parameters or perform traditional Bayesian inference. In setting where the likelihood function is not available, approximate Bayesian computation (ABC)  \cite{lintusaari2017fundamentals} offers a way to sample from an approximate posterior distribution of the parameter   $p(\parameter|\dataObs) \approx \prior(\parameter)p(\dataObs|\parameter)$  given the observed data $\dataObs$, where 
$\prior(\parameter)$ and $p(\dataObs|\parameter)$ are correspondingly the prior distribution on the parameter $\parameter$ and the likelihood function. Further we note that the mode of this approximate posterior distribution (i.e. the maximum-a-posteriori -MAP- estimate)
is also the approximate maximum likelihood estimate if we assume the prior distribution on the parameter to be uniform. 
Following this, given the observed data, we compute the MAP estimates using the ABC approximate posteriors of the parameters of the stochastic platelet deposition model.

\begin{figure*}[h!]
\caption{{\bf Approximate Bayesian computation: } Having observed data $\dataObs$ from an individual patient (the gray dot), we sample parameter values $\theta$ from the prior and generate observations through the model simulator, $\mathcal{M}(\theta)$, that are then accepted (green) or rejected (red) according to their distance from the observation measured by $\distance(\data_1,\data_2) = ||s(\data_1)-s(\data_2)) ||_2$ on the summary statistics space.}
		\label{fig:ABC}
	\begin{center}
		\includegraphics[width=\textwidth]{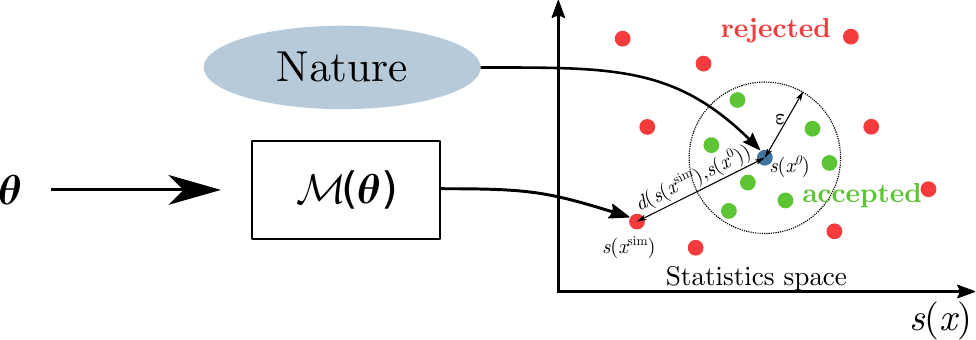}
	\end{center}
\end{figure*}

\paragraph{Approximate Bayesian computation (ABC)}
The fundamental ABC rejection sampling scheme iterates the following steps: 
\begin{enumerate}
	\item Draw $ \parameter $ from the prior $ \pi(\parameter)$.
	\item Simulate a synthetic dataset $\datasim$ from the simulator-based model $\Model(\parameter)$.
	\item Accept the parameter value $\parameter$ if $ \distance(\datasim,\dataObs) < \threshold $. Otherwise, reject $ \parameter $.
\end{enumerate}
See Figure \ref{fig:ABC} for a visualization of the above algorithm. 

Here, the metric on the dataspace $\distance(\datasim,\dataObs)$ measures the closeness between $\datasim$ and $\dataObs$. The accepted $(\parameter,\datasim)$ pairs are thus jointly sampled from a distribution proportional to $\pi(\parameter)p_{\distance,\threshold}(\dataObs|\parameter)$, where $p_{\distance, \threshold}(\dataObs|\parameter)$ is an approximation to the likelihood function $p(\dataObs|\parameter)$:

\begin{equation}\label{Eq:ABC}
p_{\distance, \threshold}(\dataObs|\parameter) = \int p(\datasim|\parameter) \mathbb{K}_{\threshold}(\distance(\datasim,\dataObs))  d\datasim, 
\end{equation}
where $\mathbb{K}_{\threshold}(\distance(\datasim,\dataObs)) $ is in this case a probability density function proportional to $\mathds{1}{(\distance(\datasim,\dataObs)<\threshold)}$\footnote{$\mathds{1}(\cdot)$ is used as an indicator function. }. Besides this choice for $\mathbb{K}_{\threshold}(\distance(\datasim,\dataObs))$, that has been exploited in several papers (for instance \cite{beaumont:2010, drovandi2011estimation, del:2012, lenormand2013adaptive}), ABC algorithms relying on other choices exist, for instance with the kernel $\mathbb{K}$ being $\exp(-\distance(\datasim,\dataObs)/\threshold )$ as in simulated-annealing ABC (SABC) \cite{Albert_2015}. More advanced algorithms than the simple rejection scheme detailed above are possible, for instance ones based on Sequential Monte Carlo \cite{del:2012, lenormand2013adaptive}, in which various parameter-data pairs are considered at a time and are evolved over several generations, while $ \threshold $ is decreased towards $0$ at each generation to improve the approximation of the likelihood function, so that we are able to approximately sample from the true posterior distribution. For inference of parameters of the platelets deposition model, here we choose the SABC algorithm, based on its suitability to high performance computing systems \cite{dutta2017abcpyhpc}. For practical implementation, SABC was run for 20 iterations generating 510 samples from the posterior distribution of the model parameters given data from each patient, keeping all other parameters fixed to the default values proposed in the Python package `ABCpy' \cite{Dutta_2017_PASC}. Next, we explain how the distance between datasets used for ABC was chosen via a discriminative summary statistics learning approach and finally how the MAP estimates (MAP) of the parameters was computed. 
\paragraph{Discriminative summary statistics learning (DSSL)}
Traditionally, distance between $\datasim$ and $\dataObs$ is defined by summing over Euclidean distances between all possible pairs composed by one simulated and one observed datapoint in the corresponding datasets. Recently, distances for ABC have also been defined through classification accuracy \cite{gutmann2017likelihood}, Kullback-Liebler divergence \cite{jiang2018approximate}, maximum mean discrepancy \cite{park2016k2} or by Wasserstein distance \cite{bernton2019approximate}, under the assumption that the datapoints in each datasets are identical and independently distributed and they are available in a large number in both $\datasim$ and $\dataObs$. We notice that in our setting we only have one datapoint in the observed dataset corresponding to the platelet deposition pattern of a patient and also due to the very expensive nature of our simulator model (simulation of one datapoint takes around 15 minutes) we can only have few datapoints in the simulated dataset. Hence, here we concentrate on the definition of distances through Euclidean distance on summary statistics extracted from the dataset when we only have one data-point in both $\datasim$ and $\dataObs$.

When the data $ \data $ is high-dimensional, a common practice in ABC literature is to define $\distance$ as Euclidean distance between a lower-dimensional summary statistics $s: \datasim \mapsto s(\datasim)$.
Reducing the data to suitably chosen summary statistics may also yield more robust inference with respect to noise in the data. Moreover, if the statistics is sufficient, then the above modification provides us with a consistent posterior approximation \cite{didelot_likelihood-free_2011}, meaning that we are still guaranteed to converge to the true posterior distribution in the limit $ \threshold \to 0 $. As sufficient summary statistics are not known for the majority of the complex models, the choice of summary statistics remains a problem \cite{csillery_approximate_2010} and they have been previously constructed using neural networks trained on a `pilot' simulated dataset from the simulator model as in neural network based semi automatic summary statistics learning (SASL) \cite{fearnhead_constructing_2012, jiang2017learning} or summary statistics learning minimizing triplet loss (TLSL) \cite{pacchiardi2020distance}. Detailed description of SASL and TLSL can be found in \hyperref[S1_App]{S1} Appendix.

These procedures are computationally expensive due to the need of the simulation of the `pilot' dataset, further the learned summary statistics using these methods are not able to discriminate between datasets from different patient types. As the main goal of the present research is to learn parameter values which are able to differentiate between different patient types, here we propose a methodology to learn such summary statistics and name it as discriminative summary statistics learning (DSSL). Learning summary statistics which is most discriminative between datasets with different labels falls under a well-developed field of research in metric-learning \cite{suarez2018tutorial}. We use Large Margin Nearest Neighbor Metric Learning (LMNN) \cite{weinberger2009distance}, one of the metric-learning approaches, which learns a Mahalanobis distance between data from any two patients $\data_1$ and $\data_2$ able to discriminate between datasets with different labels,
\begin{equation}
\label{eq:mahadist}
\distance_M(\data_1,\data_2) = \sqrt{(\data_1 - \data_2)^TM(\data_1-\data_2)}
\end{equation}
where $M$ is a $d \times d$ positive semi-definite matrix. We note that learning of the Mahalanobis distance here corresponds to learning a summary statistics of the data. It is sufficient to recall that for each positive semidefinite matrix $ M $ there exists a square matrix $L $ such that $M = L^T L$. Therefore, we can write Equation~\ref{eq:mahadist} in the following way: 
\begin{eqnarray*}
	    d_M(\data_1,\data_2) &=& \sqrt{(\data_1-\data_2)^T L^T L (\data_1-\data_2)} 
	    = \| L(\data_1-\data_2)\|_2,
\end{eqnarray*}
where $|| \cdot ||_2$ is the Euclidean distance, 
from which it is clear that the above corresponds to learning the transformation $ s: \data \mapsto s(\data) = L \data $ which is able to discriminate between different patient groups. 

\begin{figure*}[h!]
\caption{{\bf Discriminative summary statistics space:} The discriminative summary statistics space learned by DSSL, in which the patients with COPD, patients having dialysis and healthy volunteers were accurately clustered.}
	\begin{center}
		\includegraphics[width=.9\textwidth]{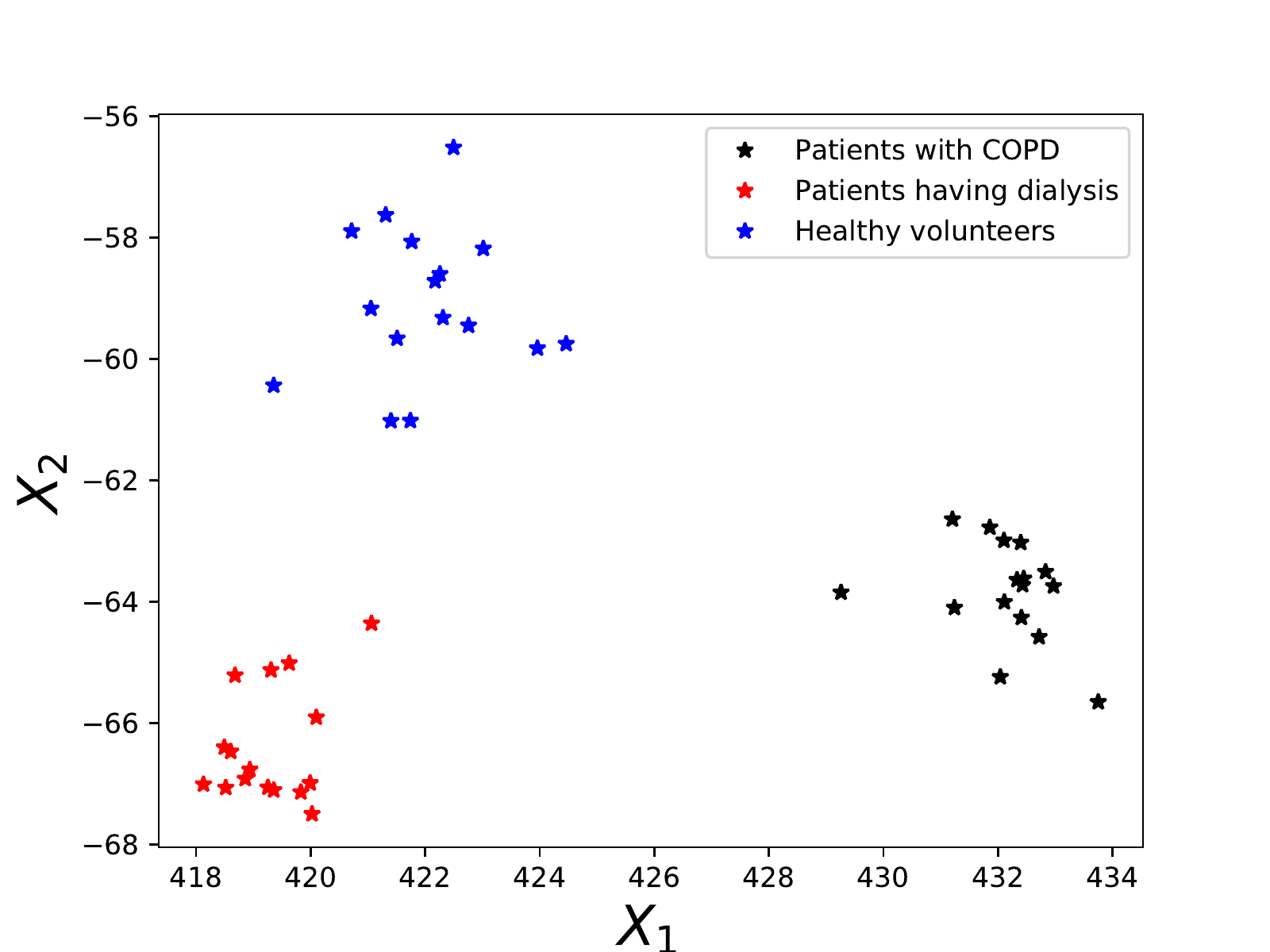}
	\end{center}
\label{fig:dssl_space}
\end{figure*}

To learn the transformation, LMNN solves the following optimization problem:
\begin{eqnarray*}\label{Eq:SDML}
\mathop{min}_{L} \sum_{i,j} \eta_{ij}||L(\data_i-\data_j)||^2 + \sum_{i,j,l}\eta_{ij}(1-y_{ij})\left[1+ ||L(\data_i-\data_j)||^2 
+  ||L(\data_i-\data_l)||^2 \right]_{+},
\end{eqnarray*}
where $\data_i$ is a data from a patient, $\data_j$ is one of that patient's k-nearest neighbors belonging to the same group of patients, and $\data_l$ are all the other data from patients of different type within the neighborhood, $\eta_{ij},y_{ij}\in\lbrace0,1\rbrace$ are indicators, $\eta_{ij}=1$ if $\data_j$ is one of the k-nearest neighbors (conditioned on being of the same type of patient as $\data_i$) of $\data_i$, $y_{ij}=0$ indicates $\data_i, \data_j$ are different types of patients, $[\cdot]_+=max(0,\cdot)$ is the Hinge loss. Intuitively, LMNN tries to learn a metric able to keeps k-nearest neighbors from the same patient group close together, while keeping patients from the other groups well separated. Further we note LMNN does not make any assumptions about the distribution of the data. To learn the projection by using LMNN, we consider the original data, its second and third order polynomial expansion and the cross products between them. Further, we manually tuned the tuning parameters for the LMNN algorithm provided in Python package `metric-learn' \cite{de2020metric}  to maximize the rand index \cite{rand1971objective} between the true patient clusters in the dataset and the clustering achieved using agglomerative hierarchical clustering with the Euclidean distance on the learned discriminative summary statistics. Euclidean distance between the learned summary statistics from the data of different patient types were able to cluster the dataset with 100\% accuracy. In Figure~\ref{fig:dssl_space} we illustrate the learned discriminative summary statistics space in which the 3 groups of patients/volunteers are accurately clustered.

\begin{figure*}[hbt!]
\caption{{\bf Comparison of predictive performance of SASL, DSSL and TLSL.} The experimental data (black solid line), 95\% predictive credibility interval (colored shaded area) and the median prediction (colored dashed line) for SASL (red), DSSL (blue) and TLSL (green) are illustrated to compare the predictive performance of SASL, DSSL and TLSL summary learning approach for $\nac$, $\sac$ and $\np$. }
\centering
		\begin{subfigure}{.32\linewidth}
			\centering
			\includegraphics[width=\linewidth]{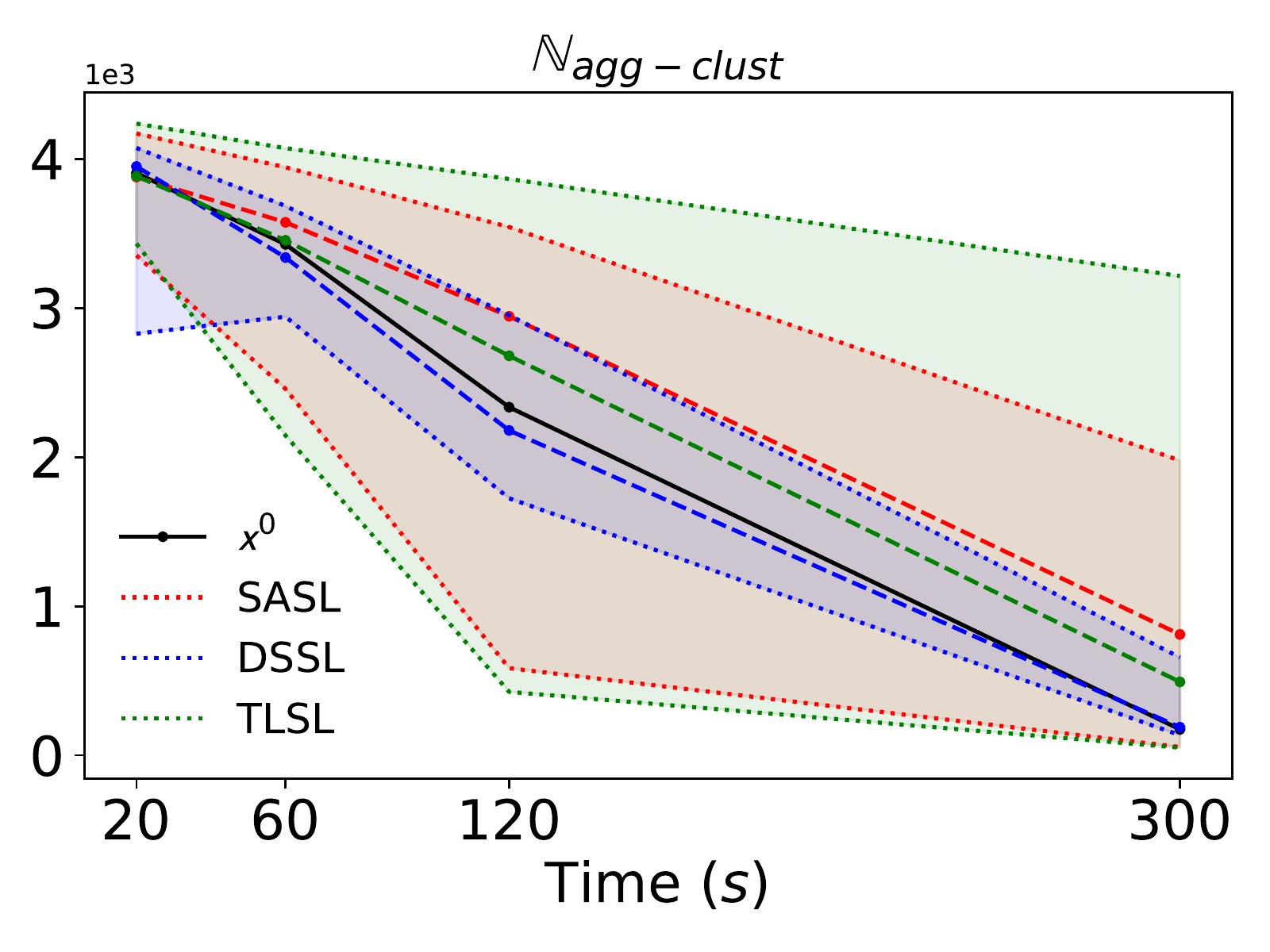}
			\caption{$\nac$}
		\end{subfigure}
		\begin{subfigure}{.32\linewidth}
			\centering
			\includegraphics[width=\linewidth]{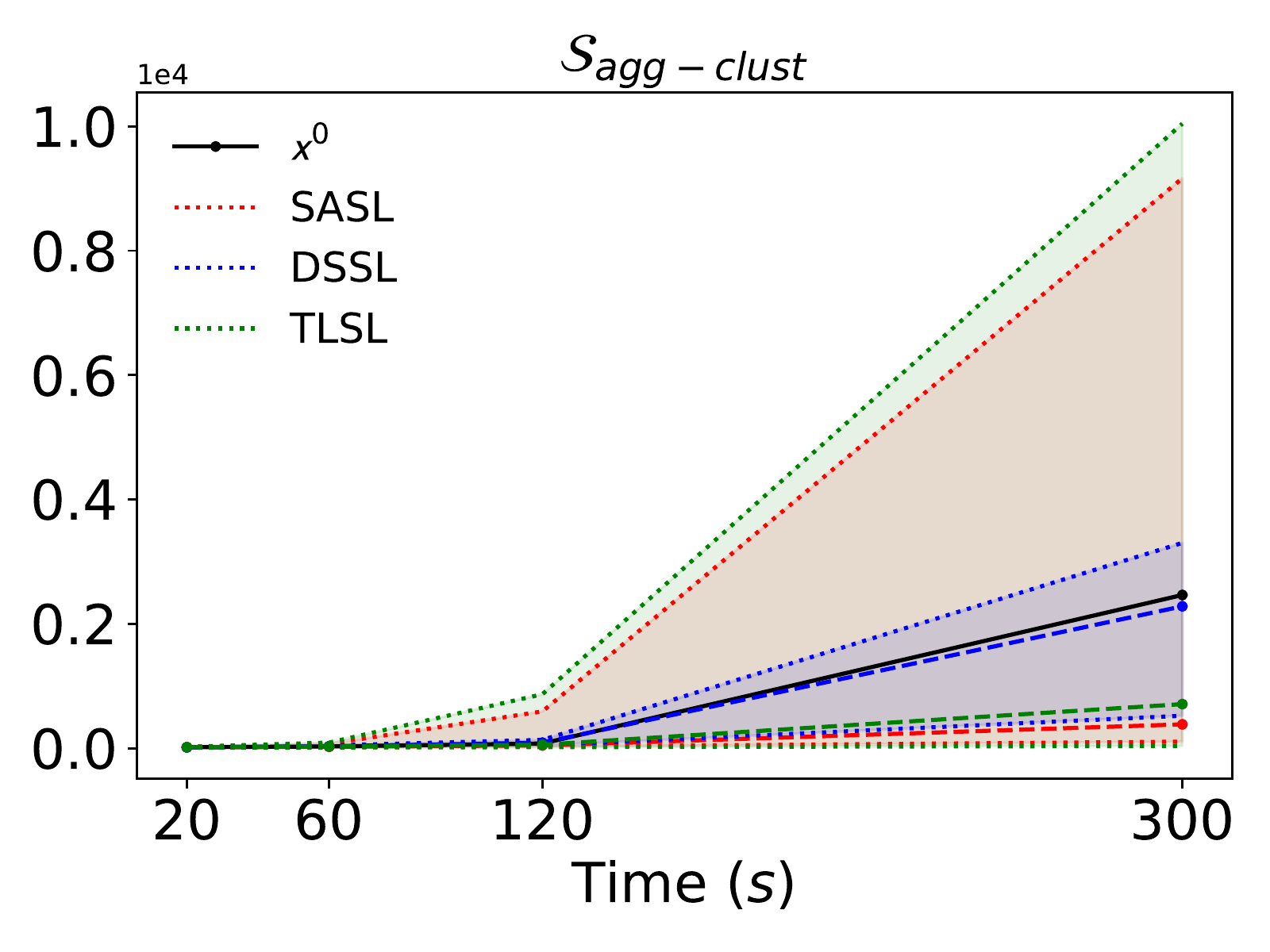}
			\caption{$\sac$}
		\end{subfigure}
		\begin{subfigure}{.32\linewidth}
			\centering
			\includegraphics[width=\linewidth]{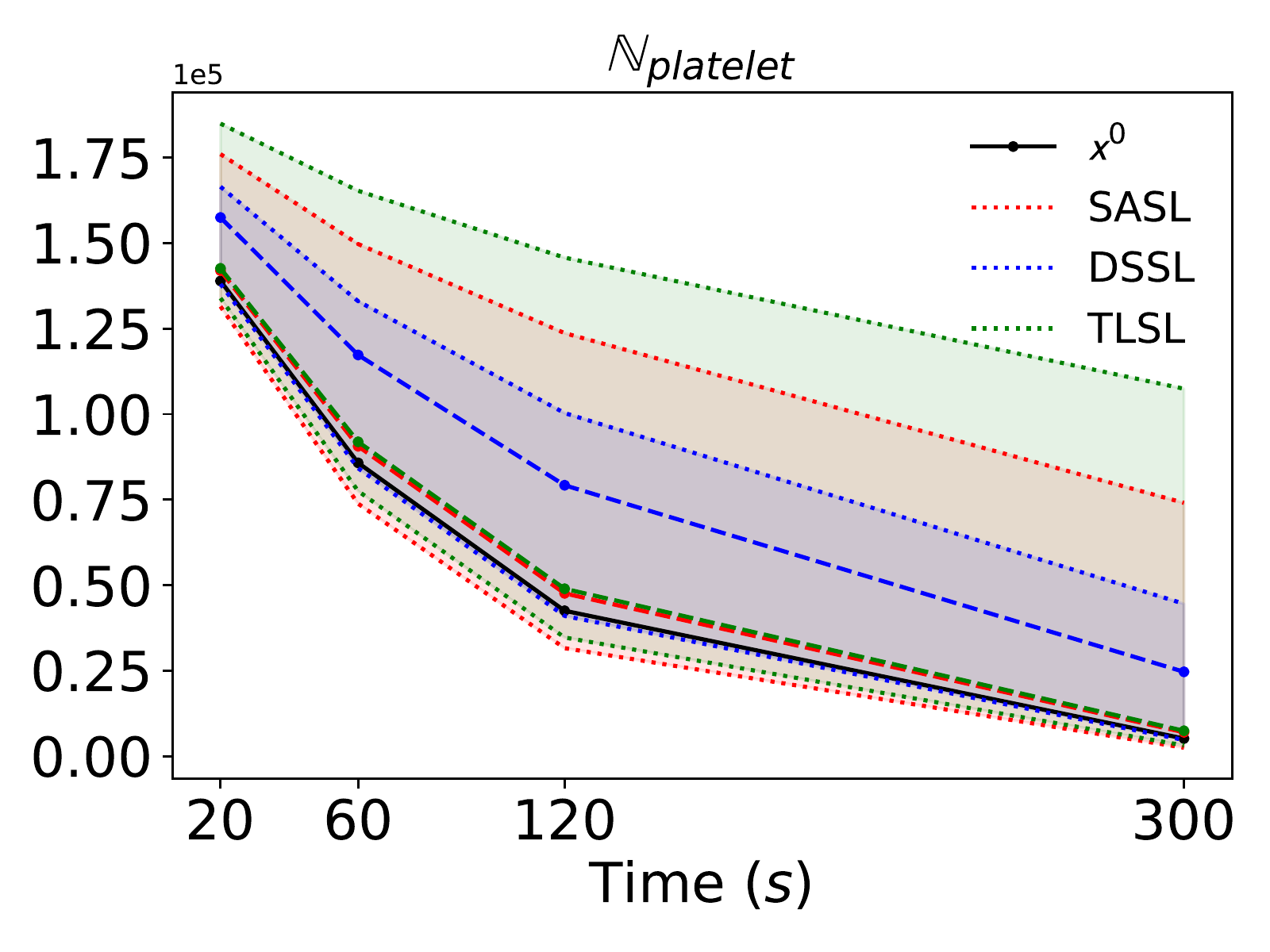}
		\caption{$\np$}
		\end{subfigure}
\label{fig:predict_comparison}
\end{figure*}

To validate our DSSL approach, we compare DSSL with SASL and TLSL by using posterior predictive checks for a experimental simulated dataset from the stochastic platelet deposition model. Further experimental details can be found in \hyperref[S1_App]{S1} Appendix. The main goal here is to analyze the degree to which the experimental data deviate from the data generated from the inferred posterior distribution of the parameters. Hence we want to generate data from the model using parameters drawn from the posterior distributions learned using the three different summary statistics learned via SASL, TLSL and DSSL. To do so, we first draw 500 parameter samples from the corresponding inferred approximate posterior distribution and simulate 500 data sets, each using a different parameter sample. This simulated dataset is considered as the predicted dataset from our inferred posterior distributions. In Figure~\ref{fig:predict_comparison}, we plot the experimental data (solid line), 95\% predictive credibility interval (shaded area) and the median prediction (dashed line) for SASL (red), DSSL (blue) and TLSL (green). The experimental data falls inside the 95\% predictive credibility interval (PCI) for all the three summary learning approach, where DSSL producing the tightest PCI. For $\nac$ and $\sac$ the median prediction is closer to the true experimental data, whereas SASL and TLSL performs better for $\np$.

\begin{table*}[hbt!]
\centering
    \caption{{\bf Comparison of predictive performance of approximate ABC posterior learned via summary statistics learned using DSSL, SASL and TLSL approach based on energy score.} Better predictive performance is measured by the smaller energy score values.}
    \label{tab:energy_score}
    \begin{tabular}{|c|c|c|c|}
    \hline
     & DSSL & SASL & TLSL \\\hline
    $\nac$ & \textbf{2.71e+02} & 1.08e+03 & 1.17e+03     \\
    $\sac$ & \textbf{3.47e+02} & 2.16e+03 & 1.99e+03     \\
    $\np$ & 7.90e+04 & 4.16e+04 & \textbf{2.46e+04}    \\
    \hline
    \end{tabular}
\end{table*}

Further, to measure the deviation of the predicted dataset from the experimental data used for inference we use a Monte Carlo estimate of the energy score, which is a strictly proper scoring rule used to measure predictive performance of probabilistic predictions \cite{gneiting2007strictly},
\begin{eqnarray}
\mbox{Energy score}= 2 \sum_{i}||\data_i - \dataObs||^{\beta}_2 - \sum_{i,j}||\data_i - \data_j||^{\beta}_2,
\end{eqnarray}
where $\data_i$ is the $i$-th data simulated using a posterior sample, $\dataObs$ is the experimental data used for inference, $\beta \in (0,2)$. In Table~\ref{tab:energy_score}, we report the energy score (fixing $\beta=1$) computed correspondingly for the three inferential schemes using summary statistics learned via SASL, TLSL and DSSL for the three observed quantities $\nac$, $\sac$ and $\np$. The values in Table~\ref{tab:energy_score} are in agreement with Figure~\ref{fig:predict_comparison}, allowing us to conclude that the proposed DSSL approach works better in predicting $\nac$ and $\sac$, whereas other two approaches perform better for the prediction of $\np$. This illustrates that the discriminative summary statistics learned by DSSL did not compromise in the overall predictive performance of the model, whereas it  also encapsulates discriminative information between the three groups of patients/volunteers.

\begin{figure*}[h!]
\caption{{\bf Discriminative projection of parameter space:} The discriminative projection of the parameter space learned from the MAP estimates, in which the patients with COPD, patients having dialysis and healthy volunteers were accurately clustered illustrating that we do not lose any discriminative information inherent in the patient dataset while learning the MAP estimates.}
	\begin{center}
		\includegraphics[width=.9\textwidth]{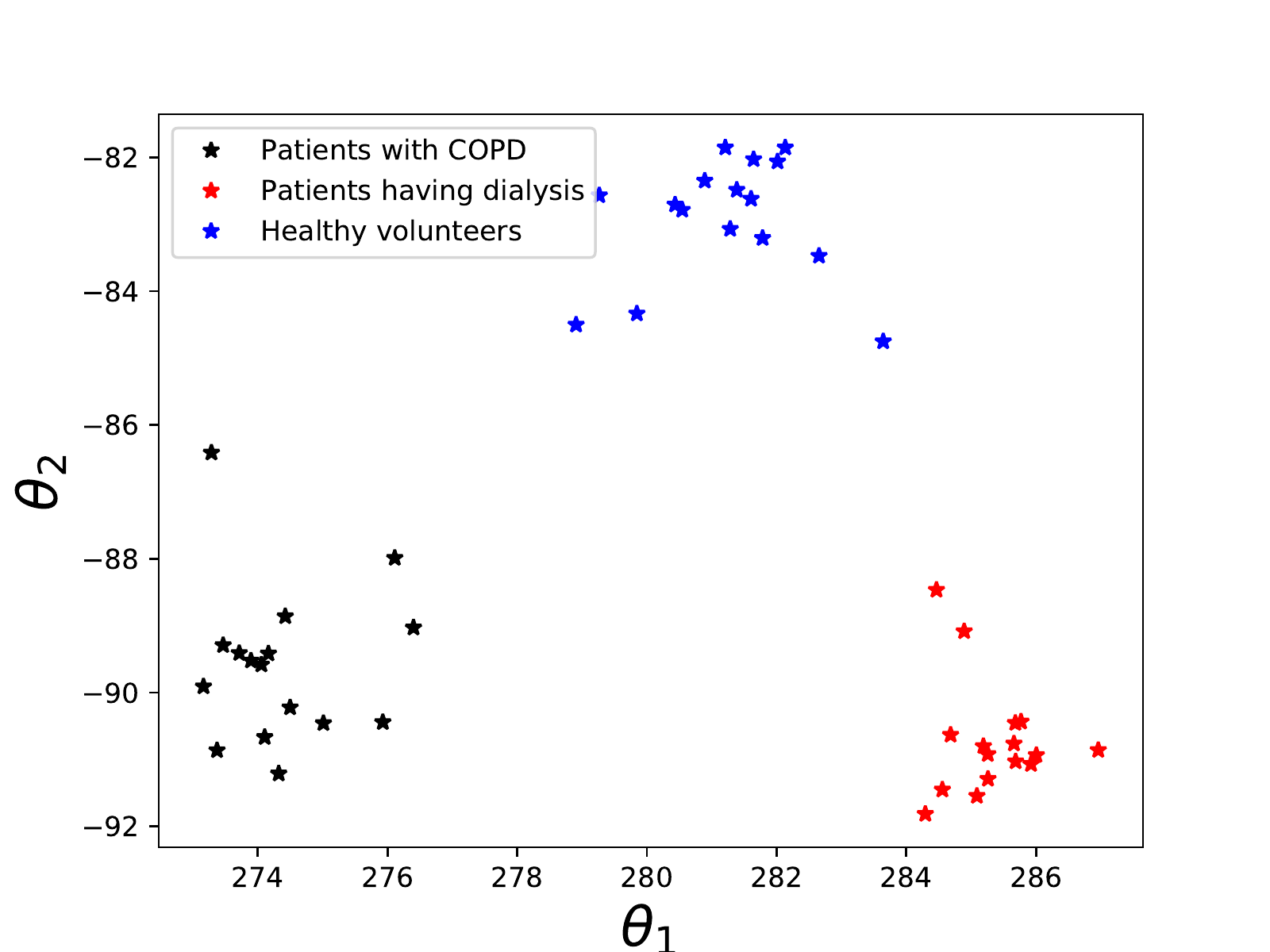}
	\end{center}
\label{fig:dparam_space}
\end{figure*}

\paragraph{Maximum a posteriori estimate (MAP)}
Given an observed dataset $\dataObs$, we want to estimate the corresponding $\parameter$. SABC inference scheme provides us with $Z$ samples $(\parameter_{i})_{i=1}^{Z}$ from the ABC approximated posterior distribution $p(\parameter|\dataObs)$ given the data for each patient. Given these samples we construct a smooth approximation of the posterior distribution (given data for a specific patient) of the parameters using Gaussian kernel density estimator with a bandwith equal to 0.45 and compute the mode of the smoothed posterior distribution using Nelder-Mead algorithm \cite{gao2012implementing} as the estimate of parameters for each specific patient. This estimate will be considered as the MAP of the parameters. The Gaussian kernel density and Nelder-Mead algorithm were used as implemented in Python package `scipy' \cite{2020SciPy-NMeth}.

In Figure~\ref{fig:dparam_space} we illustrate the discriminative projection of the parameter space $\parameter$ learnt using LMNN, by considering the MAP estimates, its second, third and fourth order polynomial expansion and the cross products between them. The tuning parameters for the LMNN algorithm were tuned as before to maximize the rand index \cite{rand1971objective} between the true patient clusters in the dataset and the clustering achieved using agglomerative hierarchical clustering with the Euclidean distance on the learned discriminative projection of parameter values from the MAP estimates. Euclidean distance between this  learned parameter projection from the MAP estimates for different patient types were able to cluster the patients with 100\% accuracy. This illustrates that we didn't lose any discriminative information inherent in the patient (volunteer) dataset by learning the MAP estimates, justifying our main claim that the values of the model parameters are precisely the information needed to assess various possible pathological conditions.

Here we further note that we could have used the discriminative summary statistics space or the discriminative projection of the parameter space correspondingly in Figure~\ref{fig:dssl_space} and \ref{fig:dparam_space}, which are not biologically meaningful, to construct the test to determine pathology in the patients with 100\% accuracy. Instead we choose here to construct the test for pathology based on the biologically meaningful parameters of our stochastic deposition model, doing so our methodology somewhat looses accuracy but gains significant interpretability.  

\paragraph{Ethics statement}Volunteers were recruited at the nephrology and pneumology units of the CHU-Charleroi, ISPPC H\^opital V\'esale in Belgium.
Written informed consent was obtained from each patient and healthy donor included in the study. The protocol of the study was in conformity with the ethical guidelines of the Helsinki Declaration of 1975 (revised in 2000) and was approved by the institution’s ethics committee (No: OM008; P17/49\_27/09).\\
Informed Consent Statement: Informed consent was obtained from all subjects involved in the study.\\

\section{Conclusion}
A numerical model of platelets deposition introduced in \cite{Chopard_2017} was illustrated to successfully predict the platelets deposition and aggregation patterns in Impact-R device. \cite{dutta2018parameter} proposed an inferential scheme based on ABC using a distance based on expert knowledge to quantify the uncertainty of the model parameters of the model presented in \cite{Chopard_2017} given data from a healthy patient. Due to the modeling of platelets reaching the bottom layers by a 1D diffusion, it was not possible to accurately calibrate the diffusion coefficient and the thickness of the boundary layers using the inferential scheme described in \cite{dutta2018parameter} and hence these quantities were manually tuned. To solve this problem, here we introduced a fully stochastic model which replaces the above parameters by the characteristic velocities of activated and non-activated platelets.  We were able to estimate these velocities in addition to the other model parameters using the inferential scheme we propose in the present manuscript. Hence the proposed methodology completely removes the need to manually tune any parameters of the numerical model of platelets deposition. 

Secondly, we adapt our inferential scheme to estimate parameters of the model which not only achieves good prediction performance but also can accommodate the discriminative information available in the dataset of different types of patients and volunteers. This is done by learning a summary statistic which is maximally discriminative between the three groups (healthy volunteers, patients with COPD and patients needing dialysis) considered and finally defining an Euclidean distance on this summary statistics space to use as distance between observed and simulated data in approximate Bayesian computation. We show that the three groups cluster accurately in this learned summary statistics space. This discriminative summary statistics was also able to get comparable predictive performance to the existing summary learning approaches when used in ABC to learn approximate posterior distribution of the model parameters given the data. 

Finally, we evaluate the maximum a posteriori of the model parameters for each of the 48 patients by computing the mode of the joint approximate posterior distribution inferred by ABC.
We notice estimated values of some of the parameters were able to distinguish between different types of patients and healthy volunteers. We would like to note here that the original learned discriminative summary statistics could be used to differentiate between different types of patients and volunteers, but those summary statistics are not interpretable in a biological sense. Hence, our main contribution lies in this ability to estimate biologically meaningful parameters which can also discriminate between different types of 
patients. This may serve as an illustration to make some of the machine learning models used in biology interpretable.

\section{Acknowledgments}
We acknowledge support from the Swiss National Supercomputing Centre (CSCS, Piz-Daint supercomputer) and the HPC Facilities of the University of Geneva (Baobab cluster). This work was funded under the embedded CSE programme of the ARCHER2 UK National Supercomputing Service (http://www.archer2.ac.uk). All the codes and dataset used for this research, can be downloaded from \href{https://github.com/eth-cscs/abcpy-models/tree/master/BiologicalScience/StochasticPlateletsDeposition}{StochasticPlateletsDepositionCode}. 

\section{Supporting information}	
\label{S1_App}
Details on Semi Automatic Summary Statistics Learning, Summary Statistics Learning by minimizing triplet loss and experimental details.
\subsection{Semi Automatic Summary Statistics Learning}
In semiautomatic summary statistics learning (SASL) schemes  \citep{fearnhead_constructing_2012, jiang2017learning}, the parameter values are regressed using some function of the corresponding simulation outputs. Namely, you assume the following model: 

\begin{equation}\label{Eq:FPNN}
\parameter = \E(\parameter| \data) + \epsilon =  f(\data) + \epsilon,
\end{equation}
where $ \epsilon $ is a 0-mean noise and $ f(\data) $ is a function of data. The authors of \cite{jiang2017learning} parametrize $ f(\cdot) $ by using a Neural Network. This regression approach was first introduced in \cite{fearnhead_constructing_2012} with a linearity assumpton on $ f $, reducing it to a simple linear regression. We focus here on the neural network formulation as this was shown to outperform the linear regression by \cite{jiang2017learning}.

In practice, we first simulate a `pilot' set of $n$ datasets $\lbrace \data_1, \ldots, \data_n \rbrace$ from $n$  parameters $\lbrace \parameter_1, \ldots, \parameter_n \rbrace$ correspondingly and then fit the statistical model given by Eq.~\eqref{Eq:FPNN} to the simulated data. Then we consider $ \statistics(\cdot) = f_\beta(\cdot) $ as the summary statistics and Euclidean distance on this summary statistics space to define the distance for the ABC inference algorithm. Following \cite{jiang2017learning}, here we use a neural network $g_w(\cdot)$ with weights $ w $ to parametrize the function $ f(\cdot) $, and that was trained by stochastic gradient descent using the loss corresponding to the regression in Eq.\eqref{Eq:FPNN}: 
\begin{equation}\label{Eq:loss_fpnn}
\frac{1}{N} \sum_{i=1}^N   ||f_\beta(\data_i) - \parameter_i||_2^2.
\end{equation}
	
In Theorem 3 of \cite{fearnhead_constructing_2012}, the authors provide a rationale for the above procedure; namely, they show that, by using $ \statistics(\dataObs) =  \E(\parameter| \dataObs)$  as summary statistics, the posterior mean of the ABC approximate posterior is the best possible estimator of the true parameter value with respect to the quadratic error loss. Of course, the posterior mean with respect to the true posterior $  \E(\parameter| \dataObs) $ is not available, and hence the regression approach was proposed. 

\subsection{Triplet Loss Summary Statistics Learning}
	
	The summary statistics learning approach minimizing triplet loss (TLSL) was first introduced in \cite{pacchiardi2020distance}, which considers the assumption that the geometry induced in data space by the Euclidean distance on the learned summary statistics ($ \statistics(\data) = g_w(\data) $ where $g_w(\cdot)$ is a neural network with weights $ w $) should be similar to the geometry in the corresponding parameter space induced by Euclidean distance ($\distance_E$). After simulating a set of $n$ datasets $\lbrace \data_1, \ldots, \data_n \rbrace$ from $n$  parameters $\lbrace \parameter_1, \ldots, \parameter_n \rbrace$ correspondingly, to learn the weights of the neural networks, here we consider the triplet \cite{schroff2015facenet} loss. The triplet loss works on three samples at a time: an anchor, a positive, that is deemed similar to the anchor, and a negative, that is on the contrary dissimilar. Essentially, the loss pushes the network to find an embedding such that the distance between the anchor and the negative is larger than the one between the anchor and the positive plus a margin, that is defined a priori. By denoting $(\data_a^{(i)}, \data_p^{(i)} , \data_n^{(i)})$ the anchor, positive and negative of the \textit{i}-th triplet, and by denoting as $ N $ the number of all possible triplets built in this way, we can write the loss in the following way: 
\begin{equation}\label{eq:triplet}
L = \frac{1}{N} \sum_i^{N} \left[|| g_w(\data_a^{(i)}) - g_w(\data_p^{(i)})||_2^2 - || g_w(\data_a^{(i)}) - g_w(\data_n^{(i)})||_2^2 + \alpha\right]_+,
\end{equation}
where $\alpha \in \R$ denotes the margin. We optimize this loss with stochastic gradient descent over the parameters of the network, by drawing random triplets. 

\subsection{Experimental Details} SASL and TLSL were trained on the same `pilot' simulated dataset containing 255 parameter and simulated data pairs. For both the SASL and TLSL the neural network is composed of 4 fully connected layers, with dimension of  input neurons and outputs being equal to the dimension of data (9) and parameter (7) correspondingly, with hidden layers of size 14, 13 and 10, batch size 16 and with ReLU non-linearity. We trained the neural network for 1000 and 2000 epochs correspondingly for SASL and TLSL. Further the margin $\alpha$ for TLSL was chosen to be 1.

\bibliographystyle{apalike}


\begin{thebibliography}{10}

\bibitem{Albert_2015}
Carlo Albert, R.~K\"unsch Hans, and Andreas Scheidegger.
\newblock A simulated annealing approach to approximate {B}ayesian
  computations.
\newblock {\em Statistics and Computing}, 25:1217--1232, 2015.

\bibitem{beaumont:2010}
Mark~A Beaumont.
\newblock Approximate bayesian computation in evolution and ecology.
\newblock {\em Annual review of ecology, evolution, and systematics},
  41:379--406, 2010.

\bibitem{benjamini1995controlling}
Yoav Benjamini and Yosef Hochberg.
\newblock Controlling the false discovery rate: a practical and powerful
  approach to multiple testing.
\newblock {\em Journal of the Royal statistical society: series B
  (Methodological)}, 57(1):289--300, 1995.

\bibitem{berger2013statistical}
James~O Berger.
\newblock {\em Statistical decision theory and Bayesian analysis}.
\newblock Springer Science \& Business Media, 2013.

\bibitem{bernton2019approximate}
Espen Bernton, Pierre~E Jacob, Mathieu Gerber, and Christian~P Robert.
\newblock Approximate {Bayesian} computation with the wasserstein distance.
\newblock {\em Journal of the Royal Statistical Society: Series B (Statistical
  Methodology)}, 81(2):235--269, 2019.

\bibitem{Xin2016}
Xin Bian, Changho Kim, and George~Em Karniadakis.
\newblock 111 years of brownian motion.
\newblock {\em Soft Matter}, 12:6331--6346, 2016.

\bibitem{breet:2010}
Nicoline~J Breet, Jochem~W van Werkum, Heleen~J Bouman, Johannes~C Kelder,
  Henk~JT Ruven, Egbert~T Bal, Vera~H Deneer, Ankie~M Harmsze, Jan~AS van~der
  Heyden, Benno~JWM Rensing, et~al.
\newblock Comparison of platelet function tests in predicting clinical outcome
  in patients undergoing coronary stent implantation.
\newblock {\em Jama}, 303(8):754--762, 2010.

\bibitem{bujak2015prognostic}
Kamil Bujak, Jaros{\l}aw Wasilewski, Tadeusz Osadnik, Sandra Jonczyk,
  Aleksandra Ko{\l}odziejska, Marek Gierlotka, and Mariusz G{a}sior.
\newblock The prognostic role of red blood cell distribution width in coronary
  artery disease: a review of the pathophysiology.
\newblock {\em Disease markers}, 2015, 2015.

\bibitem{Chopard_2015}
B.~{Chopard}, D.~{Ribeiro de Sousa}, J.~{Latt}, F.~{Dubois}, C.~{Yourassowsky},
  P.~{Van Antwerpen}, O.~{Eker}, L.~{Vanhamme}, D.~{Perez-Morga},
  G.~{Courbebaisse}, and K.~{Zouaoui Boudjeltia}.
\newblock A physical description of the adhesion and aggregation of platelets.
\newblock {\em ArXiv e-prints}, 2015.

\bibitem{chopard2017physical}
Bastien Chopard, Daniel~Ribeiro de~Sousa, Jonas L{\"a}tt, Lampros Mountrakis,
  Frank Dubois, Catherine Yourassowsky, Pierre Van~Antwerpen, Omer Eker, Luc
  Vanhamme, David Perez-Morga, et~al.
\newblock A physical description of the adhesion and aggregation of platelets.
\newblock {\em Royal Society open science}, 4(4):170219, 2017.

\bibitem{Chopard_2017}
Bastien Chopard, Daniel~Ribeiro de~Sousa, Jonas L{\"a}tt, Lampros Mountrakis,
  Frank Dubois, Catherine Yourassowsky, Pierre Van~Antwerpen, Omer Eker, Luc
  Vanhamme, David Perez-Morga, et~al.
\newblock A physical description of the adhesion and aggregation of platelets.
\newblock {\em Royal Society Open Science}, 4(4):170219, 2017.

\bibitem{csillery_approximate_2010}
Katalin Csilléry, Michael~GB Blum, Oscar~E. Gaggiotti, and Olivier François.
\newblock Approximate {Bayesian} computation ({ABC}) in practice.
\newblock {\em Trends in Ecology \& Evolution}, 25(7):410--418, 2010.

\bibitem{de2020metric}
William De~Vazelhes, CJ~Carey, Yuan Tang, Nathalie Vauquier, and Aur{\'e}lien
  Bellet.
\newblock metric-learn: Metric learning algorithms in python.
\newblock {\em Journal of Machine Learning Research}, 21(138):1--6, 2020.

\bibitem{del:2012}
Pierre Del~Moral, Arnaud Doucet, and Ajay Jasra.
\newblock An adaptive sequential monte carlo method for approximate bayesian
  computation.
\newblock {\em Statistics and Computing}, 22(5):1009--1020, 2012.

\bibitem{didelot_likelihood-free_2011}
Xavier Didelot, Richard~G. Everitt, Adam~M. Johansen, Daniel~J. Lawson, and
  {others}.
\newblock Likelihood-free estimation of model evidence.
\newblock {\em Bayesian Analysis}, 6(1):49--76, 2011.

\bibitem{drovandi2011estimation}
Christopher~C Drovandi and Anthony~N Pettitt.
\newblock Estimation of parameters for macroparasite population evolution using
  approximate bayesian computation.
\newblock {\em Biometrics}, 67(1):225--233, 2011.

\bibitem{Dutta_2017_PASC}
R~Dutta, M~Schoengens, J.P. Onnela, and Antonietta Mira.
\newblock Abcpy: A user-friendly, extensible, and parallel library for
  approximate bayesian computation.
\newblock In {\em Proceedings of the Platform for Advanced Scientific Computing
  Conference}. ACM, June 2017.

\bibitem{dutta2018parameter}
Ritabrata Dutta, Bastien Chopard, Jonas L{\"a}tt, Frank Dubois, Karim
  Zouaoui~Boudjeltia, and Antonietta Mira.
\newblock Parameter estimation of platelets deposition: Approximate bayesian
  computation with high performance computing.
\newblock {\em Frontiers in physiology}, 9:1128, 2018.

\bibitem{dutta2017abcpyhpc}
Ritabrata Dutta, Marcel Schoengens, Avinash Ummadisingu, Jukka-Pekka Onnela,
  and Antonietta Mira.
\newblock Abcpy: A high-performance computing perspective to approximate
  bayesian computation.
\newblock {\em arXiv preprint arXiv:1711.04694}, 2017.

\bibitem{fearnhead_constructing_2012}
Paul Fearnhead and Dennis Prangle.
\newblock Constructing summary statistics for approximate {Bayesian}
  computation: semi-automatic approximate {Bayesian} computation [with
  {Discussion}].
\newblock {\em Journal of the Royal Statistical Society. Series B (Statistical
  Methodology)}, 74(3):419--474, 2012.

\bibitem{gao2012implementing}
Fuchang Gao and Lixing Han.
\newblock Implementing the nelder-mead simplex algorithm with adaptive
  parameters.
\newblock {\em Computational Optimization and Applications}, 51(1):259--277,
  2012.

\bibitem{gneiting2007strictly}
Tilmann Gneiting and Adrian~E Raftery.
\newblock Strictly proper scoring rules, prediction, and estimation.
\newblock {\em Journal of the American statistical Association},
  102(477):359--378, 2007.

\bibitem{gutmann2017likelihood}
Michael~U Gutmann, Ritabrata Dutta, Samuel Kaski, and Jukka Corander.
\newblock Likelihood-free inference via classification.
\newblock {\em Statistics and Computing}, pages 1--15, 2017.

\bibitem{jiang2018approximate}
Bai Jiang.
\newblock Approximate bayesian computation with kullback-leibler divergence as
  data discrepancy.
\newblock In {\em International conference on artificial intelligence and
  statistics}, pages 1711--1721. PMLR, 2018.

\bibitem{jiang2017learning}
Bai Jiang, Tung-yu Wu, Charles Zheng, and Wing~H Wong.
\newblock Learning summary statistic for approximate bayesian computation via
  deep neural network.
\newblock {\em Statistica Sinica}, pages 1595--1618, 2017.

\bibitem{koltai2017platelet}
Katalin Koltai, Gabor Kesmarky, Gergely Feher, Antal Tibold, and Kalman Toth.
\newblock Platelet aggregometry testing: Molecular mechanisms, techniques and
  clinical implications.
\newblock {\em International journal of molecular sciences}, 18(8):1803, 2017.

\bibitem{Kotsalos2019DigitalBlood}
Christos Kotsalos, Jonas Latt, Joel Beny, and Bastien Chopard.
\newblock Digital blood in massively parallel cpu/gpu systems for the study of
  platelet transport.
\newblock {\em Interface Focus (Royal Society)}, (11):20190116, 2020.

\bibitem{kruskal1952use}
William~H Kruskal and W~Allen Wallis.
\newblock Use of ranks in one-criterion variance analysis.
\newblock {\em Journal of the American statistical Association},
  47(260):583--621, 1952.

\bibitem{lenormand2013adaptive}
Maxime Lenormand, Franck Jabot, and Guillaume Deffuant.
\newblock Adaptive approximate bayesian computation for complex models.
\newblock {\em Computational Statistics}, 28(6):2777--2796, 2013.

\bibitem{lintusaari2017fundamentals}
Jarno Lintusaari, Michael~U Gutmann, Ritabrata Dutta, Samuel Kaski, and Jukka
  Corander.
\newblock Fundamentals and recent developments in approximate bayesian
  computation.
\newblock {\em Systematic biology}, 66(1):e66--e82, 2017.

\bibitem{malerba2016platelet}
Mario Malerba, Alessia Olivini, Alessandro Radaeli, Fabio Luigi~Massimo
  Ricciardolo, and Enrico Clini.
\newblock Platelet activation and cardiovascular comorbidities in patients with
  chronic obstructive pulmonary disease.
\newblock {\em Current medical research and opinion}, 32(5):885--891, 2016.

\bibitem{mallah2020platelets}
Haneen Mallah, Somedeb Ball, Jasmine Sekhon, Kanak Parmar, and Kenneth Nugent.
\newblock Platelets in chronic obstructive pulmonary disease: An update on
  pathophysiology and implications for antiplatelet therapy.
\newblock {\em Respiratory Medicine}, page 106098, 2020.

\bibitem{mourikis2020platelet}
Philipp Mourikis, Carolin Helten, Lisa Dannenberg, Thomas Hohlfeld, Johannes
  Stegbauer, Tobias Petzold, Bodo Levkau, Tobias Zeus, Malte Kelm, and Amin
  Polzin.
\newblock Platelet reactivity in patients with chronic kidney disease and
  hemodialysis.
\newblock {\em Journal of Thrombosis and Thrombolysis}, 49(1):168--172, 2020.

\bibitem{WHO}
World~Health Organization.
\newblock http://www.who.int/mediacentre/factsheets/fs317/en/, 2015.

\bibitem{pacchiardi2020distance}
Lorenzo Pacchiardi, Pierre K{\"u}nzli, Marcel Schoengens, Bastien Chopard, and
  Ritabrata Dutta.
\newblock Distance-learning for approximate bayesian computation to model a
  volcanic eruption.
\newblock {\em Sankhya B}, pages 1--30, 2020.

\bibitem{park2016k2}
Mijung Park, Wittawat Jitkrittum, and Dino Sejdinovic.
\newblock K2-abc: Approximate bayesian computation with kernel embeddings.
\newblock In {\em Artificial intelligence and statistics}, pages 398--407.
  PMLR, 2016.

\bibitem{piagnerelli2007assessment}
Michael Piagnerelli, K~Zouaoui Boudjeltia, Dany Broh{\'e}e, Anne
  Vereerstraeten, Pietrina Piro, Jean-Louis Vincent, and Michel Vanhaeverbeek.
\newblock Assessment of erythrocyte shape by flow cytometry techniques.
\newblock {\em Journal of clinical pathology}, 60(5):549--554, 2007.

\bibitem{picker:11}
S.M. Picker.
\newblock In-vitro assessment of platelet function.
\newblock {\em Transfus Apher Sci}, 44:305--19, 2011.

\bibitem{rand1971objective}
William~M Rand.
\newblock Objective criteria for the evaluation of clustering methods.
\newblock {\em Journal of the American Statistical association},
  66(336):846--850, 1971.

\bibitem{schroff2015facenet}
Florian Schroff, Dmitry Kalenichenko, and James Philbin.
\newblock Facenet: A unified embedding for face recognition and clustering.
\newblock In {\em Proceedings of the IEEE conference on computer vision and
  pattern recognition}, pages 815--823, 2015.

\bibitem{shenkman:08}
B~Shenkman, Y~Einav, O~Salomon, D~Varon, and N~Savion.
\newblock Testing agonist-induced platelet aggregation by the impact-r [cone
  and plate (let) analyzer (cpa)].
\newblock {\em Platelets}, 19(6):440--446, 2008.

\bibitem{shreffler2020diagnostic}
Jacob Shreffler and Martin~R Huecker.
\newblock Diagnostic testing accuracy: Sensitivity, specificity, predictive
  values and likelihood ratios.
\newblock 2020.

\bibitem{suarez2018tutorial}
Juan~Luis Su{\'a}rez, Salvador Garc{\'\i}a, and Francisco Herrera.
\newblock A tutorial on distance metric learning: Mathematical foundations,
  algorithms and software.
\newblock {\em arXiv preprint arXiv:1812.05944}, 2018.

\bibitem{thijs2008mild}
A~Thijs, PW~Nanayakkara, PM~Ter~Wee, PC~Huijgens, C~Van~Guldener, and
  CD~Stehouwer.
\newblock Mild-to-moderate renal impairment is associated with platelet
  activation: a cross-sectional study.
\newblock {\em Clinical nephrology}, 70(4):325, 2008.

\bibitem{van2000asymptotic}
Aad~W Van~der Vaart.
\newblock {\em Asymptotic statistics}, volume~3.
\newblock Cambridge university press, 2000.

\bibitem{2020SciPy-NMeth}
Pauli Virtanen, Ralf Gommers, Travis~E. Oliphant, Matt Haberland, Tyler Reddy,
  David Cournapeau, Evgeni Burovski, Pearu Peterson, Warren Weckesser, Jonathan
  Bright, St{\'e}fan~J. {van der Walt}, Matthew Brett, Joshua Wilson, K.~Jarrod
  Millman, Nikolay Mayorov, Andrew R.~J. Nelson, Eric Jones, Robert Kern, Eric
  Larson, C~J Carey, {\.I}lhan Polat, Yu~Feng, Eric~W. Moore, Jake
  {VanderPlas}, Denis Laxalde, Josef Perktold, Robert Cimrman, Ian Henriksen,
  E.~A. Quintero, Charles~R. Harris, Anne~M. Archibald, Ant{\^o}nio~H. Ribeiro,
  Fabian Pedregosa, Paul {van Mulbregt}, and {SciPy 1.0 Contributors}.
\newblock {{SciPy} 1.0: Fundamental Algorithms for Scientific Computing in
  Python}.
\newblock {\em Nature Methods}, 17:261--272, 2020.

\bibitem{weinberger2009distance}
Kilian~Q Weinberger and Lawrence~K Saul.
\newblock Distance metric learning for large margin nearest neighbor
  classification.
\newblock {\em Journal of Machine Learning Research}, 10(2), 2009.

\bibitem{zouaoui2020spherization}
Karim Zouaoui~Boudjeltia, Christos Kotsalos, Daniel~Ribeiro de~Sousa, Alexandre
  Rousseau, Christophe Lelubre, Olivier Sartenaer, Michael Piagnerelli,
  J{\'e}r{\^o}me Dohet-Eraly, Frank Dubois, Nicole Tasiaux, et~al.
\newblock Spherization of red blood cells and platelet margination in copd
  patients.
\newblock {\em Annals of the New York Academy of Sciences}, 2020.

\end{thebibliography}

\end{document}